%% file: sample-acmsmall.tex
\documentclass[format=acmsmall, review=false, screen=true]{acmart}

\usepackage{booktabs} 
\usepackage{longtable}
\usepackage{multirow}
\usepackage{rotating}
\usepackage{float, lscape}
\usepackage{array}
\usepackage[ruled]{algorithm2e} 

\SetAlFnt{\small}
\SetAlCapFnt{\small}
\SetAlCapNameFnt{\small}
\SetAlCapHSkip{0pt}
\IncMargin{-\parindent}

\acmJournal{CSUR}
\acmVolume{9}
\acmNumber{4}
\acmArticle{39}
\acmYear{2017}
\acmMonth{3}
\copyrightyear{}

\setcopyright{acmlicensed}

\acmDOI{0000001.0000001}

\received{}
\received[revised]{}
\received[accepted]{}

\begin{document}
\title[A Survey on Biometric Systems and Wearables]{A Survey on Modality Characteristics, Performance Evaluation Metrics, and Security for Traditional and Wearable Biometric Systems}
\author{Aditya Sundararajan}
\author{Arif I. Sarwat}
\author{Alexander Pons}
\affiliation{%
  \institution{Florida International University}
  \streetaddress{10555 W Flagler St.}
  \city{Miami}
  \state{FL}
  \postcode{33174}
  \country{USA}}

\begin{abstract}
Biometric research is directed increasingly towards Wearable Biometric Systems (WBS) for user authentication and identification. However, prior to engaging in WBS research, how their operational dynamics and design considerations differ from those of Traditional Biometric Systems (TBS) must be understood. While the current literature is cognizant of those differences, there is no effective work that summarizes the factors where TBS and WBS differ, namely, their modality characteristics, performance, security and privacy. To bridge the gap, this paper accordingly reviews and compares the key characteristics of modalities, contrasts the metrics used to evaluate system performance, and highlights the divergence in critical vulnerabilities, attacks and defenses for TBS and WBS. It further discusses how these factors affect the design considerations for WBS, the open challenges and future directions of research in these areas. In doing so, the paper provides a big-picture overview of the important avenues of challenges and potential solutions that researchers entering the field should be aware of. Hence, this survey aims to be a starting point for researchers in comprehending the fundamental differences between TBS and WBS before understanding the core challenges associated with WBS and its design.
\end{abstract}

%
%

%
%


\thanks{This work is supported by the National Science Foundation under Grant No. 1553494. Any opinions, findings, and conclusions or recommendations expressed in this material are those of the authors and do not necessarily reflect the views of the National Science Foundation.

Author's addresses: A. Sundararajan, A.I. Sarwat and A. Pons, Department of Electrical and Computer Engineering, Florida International University (FIU), Miami, FL-33174.}

\maketitle

\renewcommand{\shortauthors}{A. Sundararajan et al.}

\input{samplebody-journals}

\end{document}

%% file: samplebody-journals.tex
\section{Introduction}
\label{sec:1}
\begin{figure}
  \includegraphics[scale=0.3]{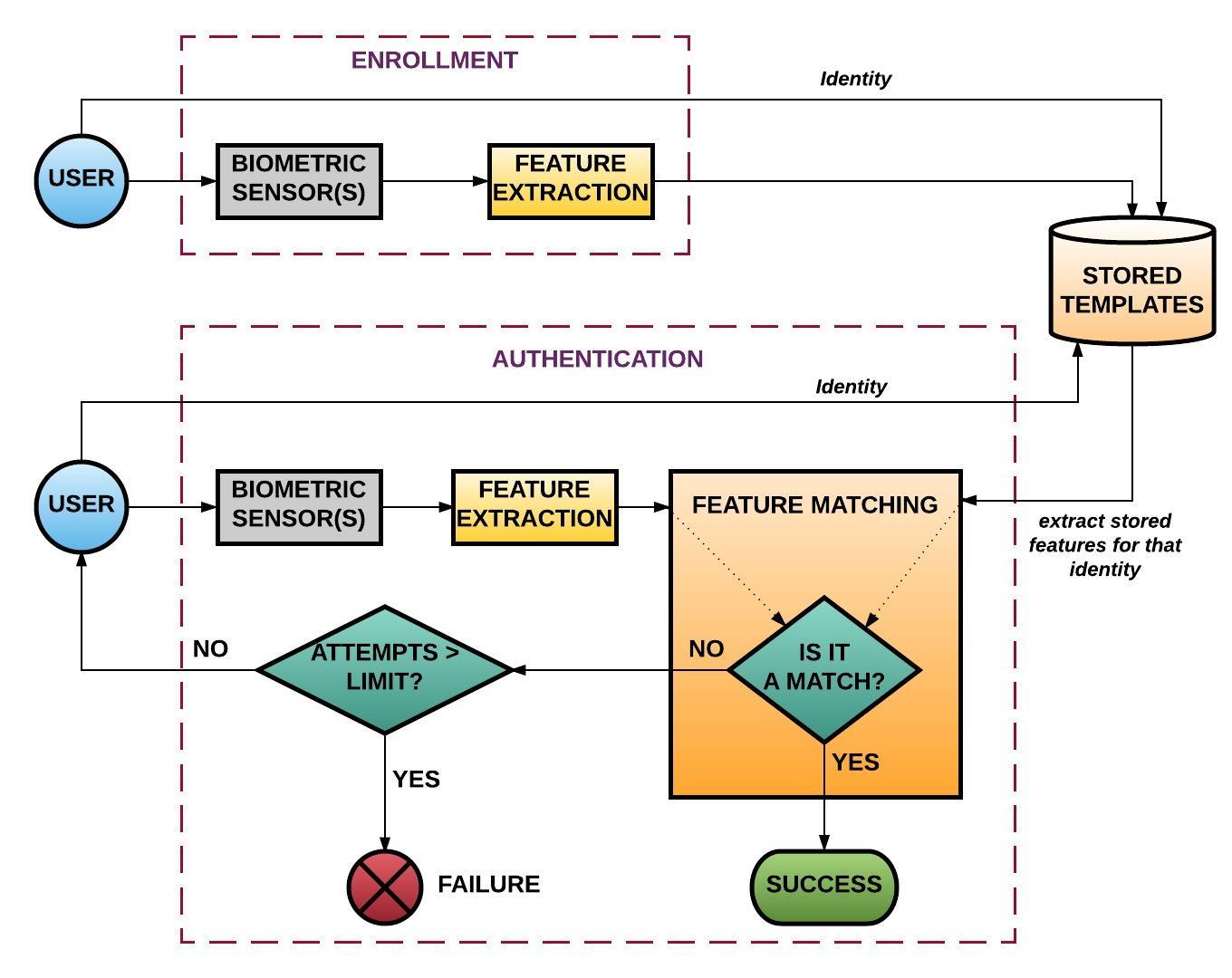}
  \caption{Block diagram for the system model of a typical TBS.}
  \label{fig:one}
\end{figure}

Biometrics is a field of science which deals with the exploitation of unique, identifiable and quantitatively measurable characteristics of humans in order to authenticate and/or identify them \cite{Roberts2006}. Over the years, pattern recognition and machine learning algorithms have found immense significance in user authentication. A hardware-software-based technology that applies such algorithms to human biometrics for authentication and, more recently, identification, fall under the class called \textit{biometric systems}. Although biometric devices are used more commonly, biometric systems has been used in this paper to emphasize that the scope of study is beyond the physical device itself, and considers communications and other applications that use the data from the individual devices. Traditionally deployed as standalone systems, the biometric systems require separate, mutually exclusive "enrollment" and "authentication" phases \cite{Angle2008}. Figure \ref{fig:one} illustrates the system model of \textit{traditional biometric systems (TBS)}. During enrollment, the user registers their traits or "modalities" as a template, created by selecting and extracting specific features from the sample recorded by sensor(s), such that it is enough to uniquely identify that user \cite{Soutar1998}. Their identity and corresponding template are then stored in the database.

At this point, security of template becomes critical, as they cannot be revoked once the database is compromised. Owing to plenty of additive noise common during measurement, hashing the template is counter-productive \cite{Sutcu2007}. There are many template protection methods like secure sketch schemes whose strength is measured by the average min-entropy of the original template given the secure sketch \cite{Sutcu2013}, fuzzy commitment scheme based on binary error-correcting codes \cite{Juels1999}, and the use of mutual information to measure dishonesty among users \cite{Boyen2005}. The Information Technology Laboratory (ITL) of the National Institute of Standards and Technology (NIST) recommends standards for biometric data exchange, system accuracy and interoperability \cite{NIST2013}.

Another class of systems are emerging, referred to in this paper as \textit{wearable biometric systems (WBS)}, which are miniaturized, mobile, flexible, comfortable, less invasive, and aesthetically pleasing. It is worth noting that WBS also come under another broader group, the wearable devices, which also include token and smartcards. However, this paper focuses only on WBS. TBS and WBS ensure user security in different ways.

\begin{itemize}
\item TBS consider each modality of the user as a separate entity while WBS consider the entire user (along with all of their individual modalities) as one \cite{Myerson2012}
\item TBS can be optionally connected to the Internet, while WBS are inherently online, exploiting the principles of Internet of Everything (IoE) \cite{ABI2014}
\item The authentication and identification processes for TBS are static and user-initiated while for WBS are dynamic and autonomous
\item While traditional modalities such as fingerprints, gait, motor, iris, and retina can be integrated into wearables using less-invasive sensors embedded in eyewear, waist-belts, etc., modalities considered invasive in TBS such as ECG, EEG, and Electromyogram (EMG) could be less invasive in WBS domain
\item Threats, attacks and defense landscapes for TBS is very different from those for WBS, considering they have different operating dynamics, characteristics and context
\item While TBS are widely used in research institutes, hospitals, libraries, airports and universities, WBS currently find their use more in healthcare and personal fitness than in security \cite{Pons2009}
\end{itemize}

\begin{figure}
  \includegraphics[scale=0.3]{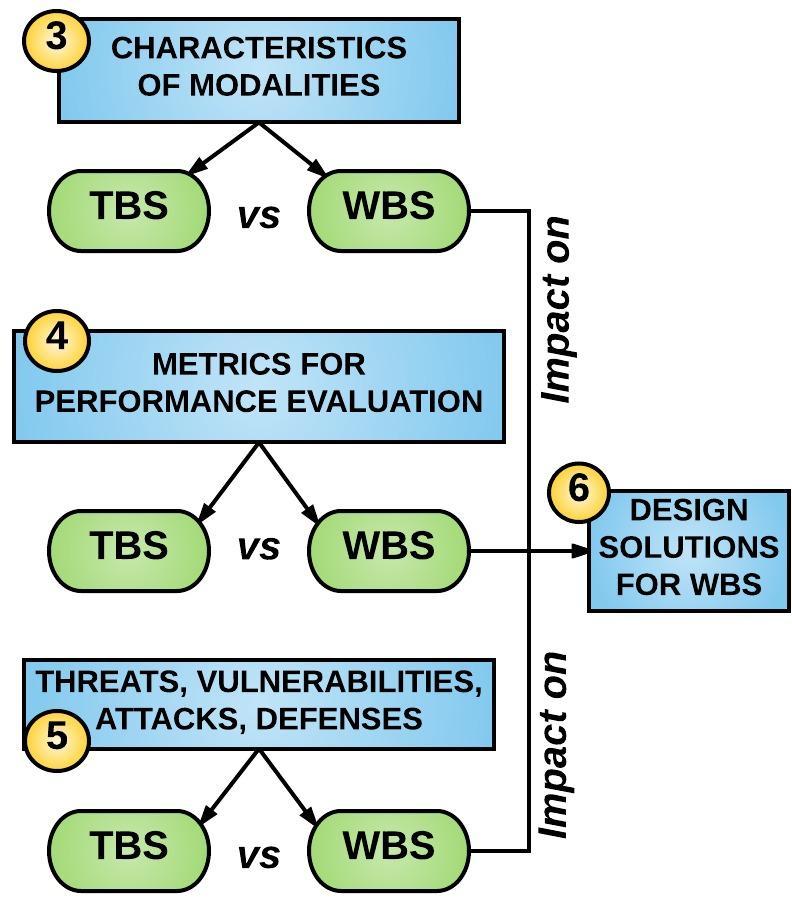}
  \caption{Flowchart showing the outline of this paper.}
  \label{fig:two}
\end{figure}

It is clear that the future of biometric systems research is geared towards WBS, especially factors like modality characteristics, performance, security and privacy, and how they impact system design and operation. However, prior to looking deeper into this area, it is important to better understand the differences between TBS and WBS highlighted above. To this end, the paper makes the following \textbf{\textit{three key contributions}}: \textbf{1)} Summarize and compare TBS and WBS in terms of the factors identified above and discuss how they contribute to the differences listed earlier; \textbf{2)} Contribute to the literature by reviewing and summarizing observations of how the modality characteristics, evaluation metrics and security impact the future design solutions for WBS; and \textbf{3)} Present important open challenges and future research directions in biometric systems that researchers should pay attention to. Hence, this survey is one of the first efforts in summarizing the research conducted beyond the widely explored TBS, and serves as a strong starting point for researchers entering/in interrelated fields.

The rest of the paper is organized as follows based on the outline illustrated by Figure \ref{fig:two}. Section \ref{sec:2} introduces WBS, their system model, and Wireless Body Area Networks (WBANs). It also summarizes various products in the market that leverage the technology. A comprehensive summary of the different key characteristics of various WBS modalities is tabulated and discussed in Section \ref{sec:3}. Section \ref{sec:4} reviews various metrics to evaluate the performance of TBS and WBS. Metrics for WBS are summarized in contrast with those of TBS. The threats, attacks and defenses for TBS and WBS are reviewed in Section \ref{sec:5}. Attacks are summarized based on whether they target the modality, technology or both to provide a cohesive organization of the literature. Defenses available against the surveyed attacks are also presented. A brief discussion on how the modality characteristics, performance evaluation metrics, security and privacy affect future design solutions for WBS is discussed in Section \ref{sec:6}. While Section \ref{sec:7} presents the open research challenges and future directions for research in the area of WBS, Section \ref{sec:8} concludes the survey by summarizing the key findings.

\section{Wearable Biometric Systems (WBS)}\label{sec:2}
Today, wearable applications quantify and personalize every action and movement undertaken by users in order to monitor their health and upgrade their lifestyles. However, in the context of security, the complexity faced by attackers increases multi-fold when such wearable applications are networked to form a WBAN, with many independent, body-friendly and reliably accurate sensors measuring and sharing data to yield a composite template. In tomorrow's fast-paced world where multiple devices are capable of interacting with each other, such dynamic and self-reliant security technologies will be required. Companies like Nymi, Google, Motorola, Apple, and Fitbit are releasing wearable devices like heart-rate monitors, physical fitness trackers, smartwatches, mobile operating systems, assisted living, elder-care, ambient-assisted living, remote automotive and home appliance control, and even navigation control \cite{Nymi2015,Bonato2005}. This implies there is a strong future in WBS security for user authentication and identification.

WBS also double up as "virtual keys" that people use to protect their sensitive assets. Their sensors are tasked with data acquisition, information communication, and decision making with little to no user intervention \cite{Latre2010}. Additionally, the nodes of WBS can be implanted, surface-mounted, or even invisible \cite{Espina2014}. While surface-mounted WBS are in the form of Smartwatch \cite{Ledger}, Fitbit \cite{Mackinlay2013}, MOTOACTV \cite{Moto2013} and Jawbone UP, implanted or embedded WBS have also started making appearances in global markets. Invisible WBS, however, are still nascent in terms of their commercial availability, although two of the leading biometric garment companies, OMSignal and Hexoskin, have launched shirts and garments designed with specially fabricated textiles made of sensors that collect and visualize body statistics in real-time \cite{Montes2015}. WBAN is discussed in Section \ref{sec:2.1}. Section \ref{sec:2.2} describes the system model for WBS, while Section \ref{sec:2.3} details the products available in the market that use WBS.

\subsection{Wireless Body Area Network (WBAN)} \label{sec:2.1}
A central notion to WBS is WBAN. Wearable sensors, implanted or surface-mounted or invisible, constitute WBAN with a typical range of $1$ to $2$ meters around the body. WBAN is derived from "Wireless Personal Area Network (WPAN)", a term coined by Zimmerman in 2001 when he studied the effect of electronics brought near human body \cite{Zimmerman1995}. With improvements in Microelectromechanical System (MEMS), data analytics and wireless communications, sensors have seen successive improvements with respect to reliability and robustness \cite{Leonov2005}. The data collected in real-time by these sensors is communicated to a sink, which could be smartphones, Personal Digital Assistants (PDAs) or Personal Computers (PCs). The collected data is fused, processed, and analyzed to offer personalization, authentication and/or identification \cite{Darwish2011}.

Wearable sensors are pervasive in nature. As identified in \cite{Cory2012}, a pervasive wearable is a colony of biometric sensors, more formally called "passive biometrics" that is unobtrusively measurable (non-invasive) and has maximal independence. They are auto-configured and do not need human intervention during enrollment. However, due to this reason, data collected by them is usually more than that collected by active biometrics like TBS. In addition, passive biometric sensors should be able to discover their peers within the same body, distinguish between those that belong to other bodies in their range, maintain secure communication, and identify the individual to whom they belong. They must also have small size and lean form-factor, and be energy-efficient and independent of positioning with respect to their target organ \cite{Filipe2015}.

\begin{figure}
  \includegraphics[scale=0.3]{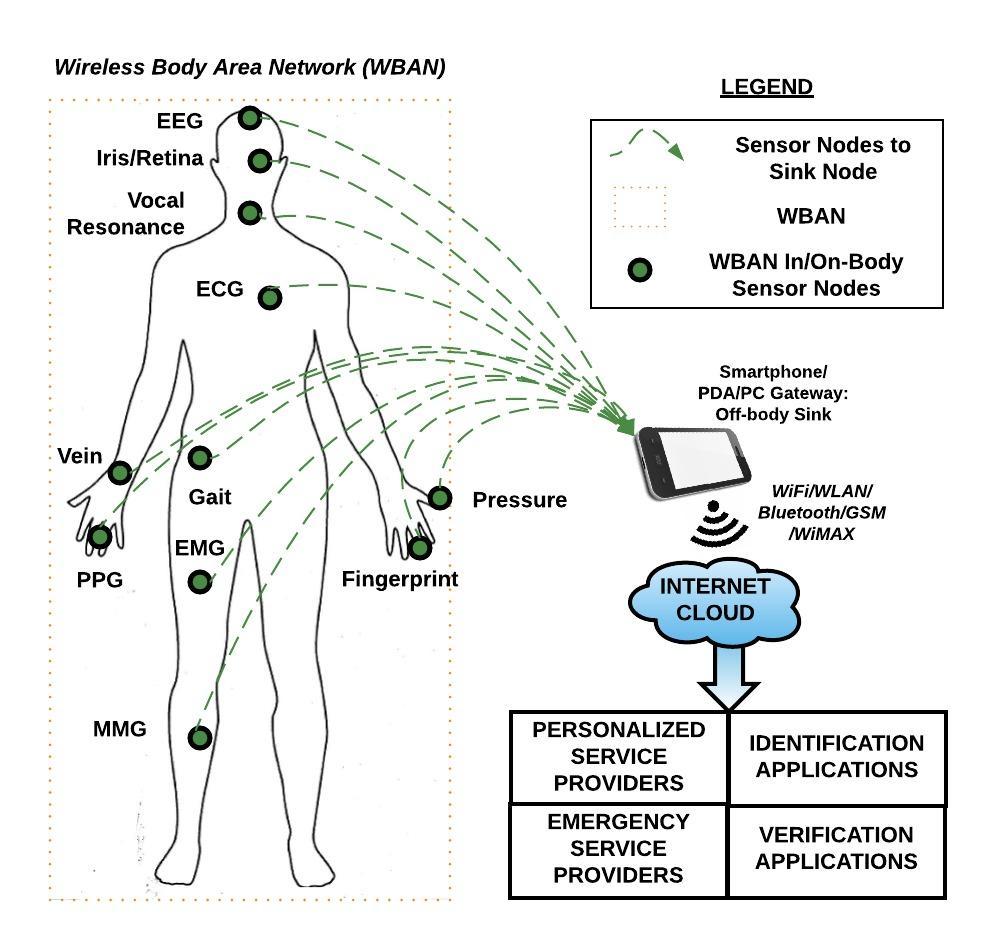}
  \caption{Block diagram of a typical WBAN with wearable sensor and sink nodes.}
  \label{fig:three}
\end{figure}

A typical WBAN, shown in Figure \ref{fig:three}, is an interconnection of multiple independent wearable sensors, each of which measure specific signals from a modality included but not limited to be one of the following: brainwaves associated with stimuli, iris structure \cite{Anderson2017}, retinal patterns \cite{Spinella2003}, vocal resonance while speaking prerecorded phrases \cite{Mitra2002, Myers2004}, skull conduction in response to audio waves propagating within the head \cite{Schneegass2016}, signals from the heart \cite{Akhter2016, Riera2011}, vein pattern on the underside of the skin, gait while walking or striding \cite{Darwish2016j, Gafurov2007}, signals generated by muscles in motion [\cite{Tarata2003}, Mechanomyograph (MMG) signals generated by muscles upon activation \cite{Mulla2014}, fingerprint patterns \cite{Chuang2014}, readings of pressure applied by fingertips when holding objects like pens, car keys, steering wheel, computer mouse, door handle, etc., patterns from signature and body odor \cite{Lorwongtragon2014}, and Photoplethysmograph (PPG) denoting the absorption of light through a body part in accordance with heartrate pulses \cite{Lee2015, Chakraborty2016}.

In WBAN, each sensor is termed a "node". The network itself can be scaled by connecting more nodes on/in the body. These nodes use wireless protocols to communicate among themselves and are coordinated by an on/in-body master "sink", forming a star topology, or communicate among themselves and independently connect to an off-body sink, forming a star-mesh hybrid topology \cite{Elias2012, Lont2014, Thotahewa2014, ICS2015}.
The protocols for communication between WBAN nodes and sink require low power communication protocols since sensors run on batteries. The IEEE 802.15.6 is considered as the primary standard that defines the Medium Access Control (MAC) architecture for in- and on-body wireless communications \cite{802156}. According to the standard, every node and sink has a Physical (PHY) layer (constituting a narrowband, ultra-wideband and human body communication PHY layers) and a MAC sublayer \cite{Li2012}. The MAC Service Data Units (MSDUs) are transferred from the MAC client layer to the MAC sublayer through the MAC Service Access Point (SAP). The MAC frames are then transferred to the PHY layer through the PHY SAP. The reverse happens at the receiving end, which would be a sink in a star topology, or a node/sink in a star-mesh hybrid topology. Network and Application layers provide configuration, routing and management, and functional requirements, respectively \cite{802154n, 802154n2}.

Standard protocols for WBAN MAC communication such as ZigBee-MAC, Baseline-MAC (BMAC), SPARE-MAC, T-MAC and D-MAC have been widely used to implement WBANs, which account for traffic resolution, collision contingencies, energy-saving, auto-configuration, throughput, and delay \cite{Sarika2013}. MAC layers employ Time Division Multiple Access (TDMA) and its hybrid variants such as $\lambda$-MAC, and A-MAC \cite{Parker2010, Rhee2008}. Carrier Sense Multiple Access with Collision Avoidance (CSMA/CA) technologies are also used for energy-efficient communication through sleep scheduling, high channel utilization and low latency \cite{Singh2013}. A Low-Energy Adaptive Clustering Hierarchy (LEACH) routing protocol was presented to uniformly distribute energy load among WBAN nodes. This method is a significant attempt towards reducing energy dissipation through randomization, and also supports scalability, adaptiveness, and robustness \cite{Wendi2000}. WBAN communication should be cognizant of electromagnetic wave diffraction due to the continuous absorption and reflection as waves pass through tissues and cells which predominantly contain water. If the sink is located off-body, normal human body movements and posture when stationary would determine the quality of information transmitted. This in-turn determines channel conditions and factors like latency and throughput, which indirectly affect energy consumption \cite{Filipe2015}.

\subsection{System Model of WBS} \label{sec:2.2}

\begin{figure}
  \includegraphics[scale=0.3]{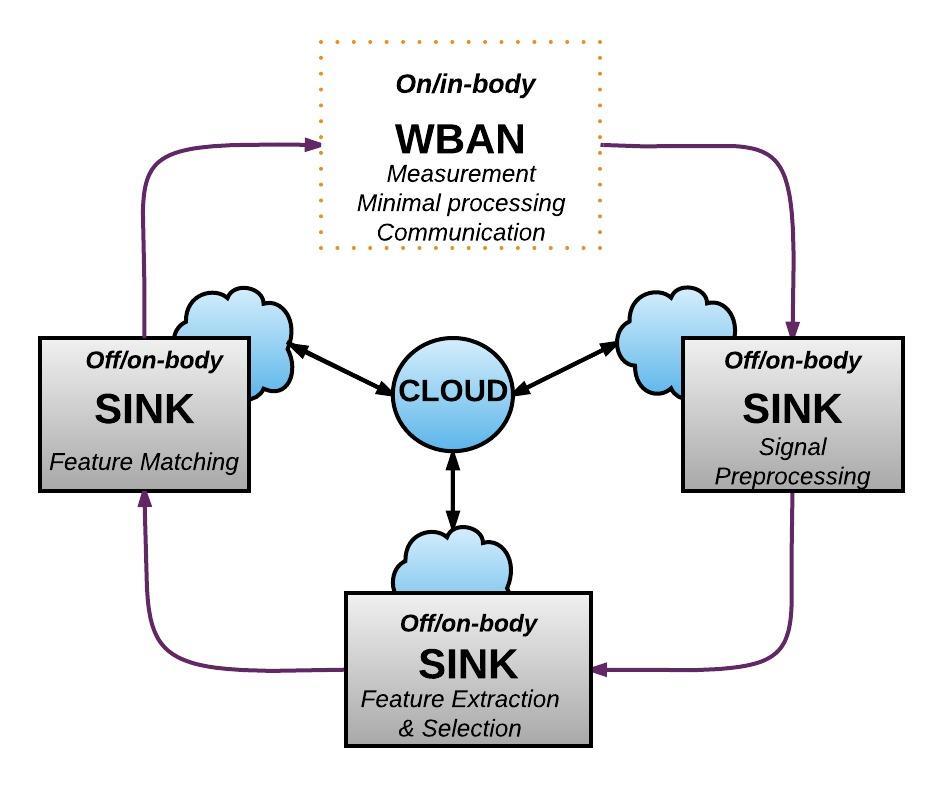}
  \caption{System model of a typical Wearable Biometric System.}
  \label{fig:four}
\end{figure}

The system model of WBS is shown in Figure \ref{fig:four} is a cyclical "continuous/online" process, where wearables authenticate or identify users not on-demand but continuously. WBS comprise the WBAN represented in Figure \ref{fig:three}, the sink, and storage, which may or may not belong to the sink. WBAN is bounded by constraints related to memory for storage and computation, network bandwidth for communication, and energy consumption, among other key factors like interoperability, form factor and bio-friendliness \cite{Blasco2016}. Some WBANs are also capable of doing minimal signal processing. Sink can be functionally decomposed into a signal preprocessing unit, feature extraction and selection unit, feature matching unit, and optionally a data storage and management unit. In some cases like Fitbit or Google Glass, the sensor and sink nodes are integrated into one device, while the storage and maintenance is pushed to cloud. Signal preprocessing involves operations such as smoothing, interpolation, and normalization. Feature extraction and selection employ standard methods like Principal or Independent Component Analysis (PCA/ICA), Linear Discriminant Analysis (LDA), logistic regression for minimal hardware complexity, and Short Time Fourier Transform (STFT) for minimal memory and computation complexity \cite{Page2015,Smital2016,Bianchi2006}. Feature matching uses one-class models for authentication and multi-class models for identification based on kNN, Support Vector Machines (SVMs), Bayesian Networks and Naïve Bayes, and Artificial Neural Networks (ANNs) for supervised learning; and Hidden Markov Models (HMMs) and Gaussian Mixture Models (GMMs) for unsupervised learning. Sink-level storage lowers vulnerability to external threats, but limits computational capabilities and increases form factor.

\subsection{Products In the Market Using WBS} \label{sec:2.3}
It was estimated that the market for WBS in sports and healthcare would reach at least $170$ million devices by the end of 2017 \cite{Valencell2013}. WBS use both physical as well as behavioral modalities, defined later in Section \ref{sec:3}, where the former are more static and the latter more dynamic. The behavioral modality-based systems employ hand-eye coordinations, user interactions, pressure, tremors, and other finer movements in addition to the modality itself to capture the data needed for authentication. This sophistication can be more easily achieved by WBS than TBS given their mobility and proximity to the human body. Behavioral modalities have also been used in continuous authentication, sometimes with randomized selection of features from a pool to base the analysis on. For example, EEG-based sensors, which measure stimulus-specific brainwaves to authenticate users, have been of recent interest to the research community, and their application to WBS has attracted decent attention. Below are a few other examples of emerging and recently available products in the market that leverage WBS and mostly behavioral modalities:
\begin{enumerate}
\item \textbf{EEG:} It is difficult with today's technological advancement to consider EEG as one of the unobtrusively measurable biometrics, since it needs the user to be stationary and calm, and to wear a head-cap comprising electrodes. However, products developed by Muse and Kokoon can be viewed as a productive step towards realizing EEG-based WBS in the future
  \begin{itemize}
  \item Muse has developed an application which measures EEG from the user's temples to which its sensors are attached \cite{InteraXon2014}
  \item  Kokoon provides a set of EEG headphones which measure user's mental state or mood using brainwaves emitted, and accordingly play music \cite{Prindle2015}
  \end{itemize}
\item \textbf{ECG:} It measures signals emitted by one's heart that are unique to the individual and is more applicable for WBS \cite{Silva2015}. Unlike EEG, ECG can be unobtrusively measured using just an electrode attached to the skin
  \begin{itemize}
  \item Bionym is a company that has developed the Nymi wristband to use heartbeat profiles for uniquely identifying users and improving their lifestyle by personalizing their living environment \cite{Nymi2015}
  \end{itemize}
\item \textbf{Electromyogram (EMG):} It being leveraged by sports clothing companies, Athos and Myontech, to design sportswear more conducive for superior athletic performance \cite{Kunju2009}
	\begin{itemize}
	\item MyoWare has developed a hardware that facilitates the control of video games, robotic movements, and actions of prosthetics based on movement profiles exhibited by the user's motor neurons \cite{Berglin2011}
	\end{itemize}
\item \textbf{Vocal resonance:} It refers to the voice of a user measured as the vocal sound waves propagate through their body as against traditional voice recognition systems that use the waves captured over air \cite{Cory2014rep}. Placing contact microphones on the neck of the user, the system can unobtrusively measure and model their vocal resonance
\item \textbf{TIAX LLC} has designed and developed wearable sensors and algorithms which can derive vital signs and bio-signatures
	\begin{itemize}
	\item To facilitate seamless coordination and communication with minimal user intervention, the company has adopted sensor-fusion methods that combine different streams of physiological data such as EEG and ECG, fingerprints, and EMG response
    \item The application of learning algorithms for discovering and exploiting unprecedented patterns is also being conducted \cite{Hill2015}
	\end{itemize}
\item In December 2016, \textbf{Valencell} and \textbf{STMicroelectronics} launched a Scalable Development Kit (SDK) for biometric wearable and Internet of Things (IoT) sensor platform
    \begin{itemize}
    \item It uses SensorTile of the STMicroelectronics, an IoT sensor module with one STM32L4 microcontroller, a Bluetooth Low Energy chipset, a host of MEMS sensors including accelerometer, magnetometer, pressure and temperature sensors that are both highly accurate as well as energy-efficient, and a digital MEMS microscope \cite{Lee2016}
    \item It is combined with Valencell's Benchmark biometric sensor system platform, and together enable the development and support of advanced wearable applications, including biometric authentication and identification
    \item Valencell has also introduced technologies to measure heart-rate using forearm, and earbud-based sensors for evaluating energy expenditure, net calories burned and maximum oxygen consumption, and magnetic sensors for measuring cadence \cite{Eschbach, Eschbachearbud, Leboeuf2014}
    \end{itemize}
\item \textbf{Skull Conduction:} An integrated bone conduction speaker was designed and proposed at the CHI Conference on Human Factors in Computing Systems in May, 2016, where the authors introduced an embedded wearable, SkullConduct, integrated with wearable computers like Google Glass \cite{Schneegass2016}
	\begin{itemize}
	\item A 1-Nearest Neighbor (1NN) classifier was used in conjunction with Mel Frequency Cepstral Coefficient (MFCC) features to analyze and interpret unique frequency response registered by the skull in response to the propagation of soundwaves within its bone structure
    \item It was touted to be stable, universal, collectible, non-invasive and robust. It was proposed both as a verification as well as an identification system
	\end{itemize}
\item \textbf{Signature:} The work focused on using it as a form of identification, where an unknown signature was used as an input to verify if a claimed identity was forged \cite{Nassi2016}
\item \textbf{Touch:} This work leveraged touch-based behavioral modality and combined it with smart eyewear to conduct continuous user authentication
	\begin{itemize}
	\item To reduce the invasiveness in authentication, the authors in this work proposed an eyewear that captures user interactions such as single tap, forward, downward and backward swipes, and backward two-finger swipes to perform continuous, online authentication of the user~\cite{peng2017}
    \item This has been touted to achieve a detection rate of $99\%$ and a false alarm rate of $0.5\%$ under an equal probability of occurrence of all six perceived user events
	\end{itemize}
\end{enumerate}
 All of the above applications more often than not consider the use of individual biometrics as wearables, in which case certain significant challenges fail to come to the forefront. However, when a colony of sensors are networked together to constitute a WBAN, security concerns become more pronounced, as will be explored in Section \ref{sec:4.2}.

\textbf{Key Takeaway Points:} The following are the key takeaway points from this section:
\begin{enumerate}
\item WBAN is a central component of WBS and has a typical range of $1$ to $2$ meters around the user's body and comprises multiple sensors that can be implanted, embedded or invisible, each dedicated to unobtrusively and independently measure specific modalities. Such wearables are called passive biometric systems
\item The nodes of WBAN use wireless protocols to communicate among themselves and are coordinated by an on, in or off-body sink
\item WBAN communication protocols must account for traffic resolution, collision contingencies, energy-saving, auto-configuration, throughput and delay
\item WBS system model comprises WBAN, and units for signal preprocessing, feature extraction and selection, feature matching, and cloud-powered data storage and management
\item There exist multiple WBS products in the market that utilize different modalities, either in isolation or combination, such as EEG, EMG, vocal resonance, fingerprints, skull conduction, signature, PPG, heart-rate, and touch
\item Some of the forerunner companies that have released WBS products include Google, Fitbit, Nymi, Myontech, Valencell, Apple, Motorola, Muse, Kokoon, TIAX LLC, and MyoWare
\end{enumerate}

\section{Key Characteristics of WBS Modalities}\label{sec:3}
Modalities can be categorized in various ways depending on the criteria. For instance, based on their type, they can be classified as physiological (iris, hand, retina, fingerprint, DNA, earlobe, Electrocardiogram (ECG) and odor among others), cognitive (Electroencephalogram [EEG]) and behavioral (signature, keystroke, voice and gait among others) \cite{Babich2012,Ali2016}. All of these modalities comprise certain characteristics like \textit{universal}- should be possessed by every person, \textit{unique}- should distinguish any two individuals, \textit{permanent}- should not be drastically affected by age or fatigue, \textit{collectible}- should easily be acquired by non-invasive means, \textit{acceptable}- should be approved by the public for widespread use, and \textit{circumventive}- should not allow adversaries to easily bypass it \cite{Nazmeen2014,Cory2014}. The modality characteristics for TBS was summarized in an earlier work \cite{Aditya2014}.

Ideally, the modalities used by WBS, in addition to the above six characteristics for TBS, must have the following characteristics: \textbf{1) Security:} overall confidence that users can invest in the modality against the loss of confidentiality, integrity or availability, \textbf{2) Reliability:} extent to which the modality can be trustworthy, \textbf{3) Mobility:} extent to which the modality's use does not get impacted by constant or intermittent motion with respect to sink, \textbf{4) Variability:} extent to which the above characteristics change with respect to time, \textbf{5) Interference:} extent to which the above characteristics get adversely affected by the presence of physical or electromagnetic impediments, and \textbf{6) Invasive:} extent to which measurement of the modality creates user discomfort.

The constraints of WBANs like communication latency, throughput and amount of energy consumed, in addition to security and resiliency, can be regarded as factors to determine the characteristics. Table \ref{tab:one} summarizes the different WBS modality characteristics, where each fares differently in satisfying the identified factors. The extent to which each modality contributes to enhance favorable factors (like throughput, goodput, security, resiliency) and reduce unfavorable factors (like energy consumed, latency, size) is used to qualitatively rank their suitability to a particular characteristic: $H$ stands for High suitability, $M$ for Medium and $L$ for Low. As explained earlier in Section \ref{sec:1}, the modalities, irrespective of being applied to TBS or WBS, must be unique, universal, permanent, collectible, acceptable and circumventive in nature. 

The Table shows how the modalities shown in Figure \ref{fig:two} fare with the key characteristics identified for WBS. It can be inferred that while some modalities fare poorer for certain characteristics when implemented in TBS, they fare better on WBS. For example, iris or retinal pattern might be cumbersome to measure when implemented in TBS since the users are required to place their eye in line and close to the camera. However, the same when implemented in WBS is very easy to measure, considering users can use smart eye-wear like Google Glass to efficiently measure iris structures and even retinal patterns in a less invasive manner.

\textbf{Key Takeaway Points:} The following are the key takeaway points from this section:
\begin{enumerate}
\item Biometric modalities must have six fundamental characteristics. They should be universal, permanent, collectible, unique, acceptable, and circumventive
\item The modalities used by WBS have additional characteristics: security, reliability, mobility, variability, interference and invasiveness
\item \label{item3} The constraints of WBANs like communication latency, throughput/goodput, and amount of energy consumed, can be regarded as primary factors that determine the system's operational dynamics
\item While some modalities fare poorer for certain characteristics when implemented in TBS, they fare better on WBS as they could be easier to measure
\item Modalities that have a "Low" suitability for a particular characteristic in TBS domain might possess a "High" suitability for the same characteristic in WBS domain, considering the two systems have different operational dynamics
\item It can, hence, be understood that the operational dynamics identified in point number (\ref{item3}) significantly shape the extent to which the modalities contribute favorably to their different characteristics, thereby showcasing a strong dependency between the two
\item Between physiological and behavioral modalities, further explained in Section \ref{sec:3}, the latter have more dynamism, and hence are better suited for continuous authentication applications that would also not increase the invasiveness or jeopardize the privacy of users
\end{enumerate}

\begin{landscape}
    \begin{table}
            \caption{Summary of Key Characteristics of Different Modalities for Wearable Biometric Systems}
        \label{tab:one}
        \centering
        \small
        \def\arraystretch{2}
        \setlength{\tabcolsep}{3pt}
        \begin{tabular} {|p{1.8cm}|p{1cm}|p{1cm}|p{1cm}|p{1cm}|p{1cm}|p{1cm}|p{1cm}|p{1cm}|p{1cm}|p{1cm}|p{1cm}|p{1cm}|p{1cm}|p{1cm}|p{1cm}|}
        % ...
        \toprule
 & \centering \textbf{Vein} & \centering \textbf{Odor} & \centering \textbf{Sign ature} & \centering \textbf{Finger print} & \centering \textbf{Retina} & \centering \textbf{Iris} & \centering \textbf{Face} & \centering \textbf{Gait} & \centering \textbf{Pres sure}& \centering \textbf{EMG/ MMG} & \centering \textbf{Voice} & \centering \textbf{ECG} & \centering \textbf{EEG} & \centering \textbf{PPG} & \multicolumn{1}{c}{\textbf{Skull}}\\\hline
\bottomrule
\centering Unique & \centering H & \centering M & \centering M & \centering H & \centering H & \centering H & \centering M & \centering M & \centering M & \centering H & \centering M & \centering H & \centering H & \centering H & \multicolumn{1}{c}{H} \\\hline
\centering Universal & \centering H & \centering M & \centering L & \centering M & \centering H & \centering H & \centering H & \centering M & \centering M & \centering H & \centering M & \centering H & \centering H & \centering H & \multicolumn{1}{c}{H} \\\hline
\centering Permanent & \centering H & \centering M & \centering M & \centering H & \centering H & \centering H & \centering M & \centering M & \centering M & \centering H & \centering M & \centering H & \centering H & \centering H & \multicolumn{1}{c}{H} \\\hline
\centering Collectible & \centering H & \centering M & \centering M & \centering H & \centering H & \centering H & \centering M & \centering M & \centering M & \centering H & \centering M & \centering H & \centering H & \centering H & \multicolumn{1}{c}{H} \\\hline
\centering Acceptable & \centering H & \centering M & \centering M & \centering H & \centering H & \centering H & \centering M & \centering M & \centering M & \centering H & \centering M & \centering H & \centering H & \centering H & \multicolumn{1}{c}{H} \\\hline
\centering Circumventive & \centering H & \centering M & \centering M & \centering H & \centering H & \centering H & \centering M & \centering M & \centering M & \centering H & \centering M & \centering H & \centering H & \centering H & \multicolumn{1}{c}{H} \\\hline
\centering Secure & \centering H & \centering M & \centering M & \centering H & \centering H & \centering H & \centering M & \centering M & \centering M & \centering H & \centering M & \centering H & \centering H & \centering H & \multicolumn{1}{c}{H} \\\hline
\centering Reliable & \centering H & \centering M & \centering M & \centering H & \centering H & \centering H & \centering M & \centering M & \centering M & \centering H & \centering M & \centering H & \centering H & \centering H & \multicolumn{1}{c}{H} \\\hline
\centering Mobile & \centering H & \centering M & \centering M & \centering H & \centering H & \centering H & \centering M & \centering M & \centering M & \centering H & \centering M & \centering H & \centering H & \centering H & \multicolumn{1}{c}{H} \\\hline
\centering Variable & \centering H & \centering M & \centering M & \centering H & \centering H & \centering H & \centering M & \centering M & \centering M & \centering H & \centering M & \centering H & \centering H & \centering H & \multicolumn{1}{c}{H} \\\hline
\centering Interferential & \centering H & \centering M & \centering M & \centering H & \centering H & \centering H & \centering M & \centering M & \centering M & \centering H & \centering M & \centering H & \centering H & \centering H & \multicolumn{1}{c}{H} \\\hline
\centering Invasive & \centering H & \centering M & \centering M & \centering H & \centering H & \centering H & \centering M & \centering M &\centering  M & \centering H & \centering M & \centering H & \centering H & \centering H & \multicolumn{1}{c}{H}\\\hline
        \end{tabular}
    \end{table}
\end{landscape}

\section{Performance Evaluation Metrics for TBS and WBS}\label{sec:4}
Although modalities typically satisfy most characteristics described in Section \ref{sec:2.3}, they focus on the modality itself but not on the products that use the modality. Performance is more applicable for products that use biometrics rather than the biometric itself. The characteristics summarized in Table \ref{tab:one} also affect the performance of such technologies. With the integration of high-end data fusion, analytics, processing, and personalization in order to securely authenticate and/or identify users, performance can be considered an important part of evaluating biometric systems.

\vspace{-0.3cm}
\subsection{Performance Evaluation Metrics for TBS}\label{sec:3.1}
Evaluating the performance of TBS has been a widely studied topic in literature \cite{Elabed2012,Precise2014,Giot2013,Dmitry2009}. A comprehensive summary of different performance evaluation metrics and charts used for TBS was described and summarized in \cite{Poh2012}. The work documents that in accordance with the ISO/IEC standard 19795 Parts 1 and 2 for biometric system performance evaluation, multiple metrics are used for verification, such as:
\begin{itemize}
\item False Match Rate ($FMR$): probability of the technology wrongly authenticating an individual claiming to belong to his correct identity
\item False Non-Match Rate ($FNMR$): probability of the system wrongly rejecting an individual claiming to belong to his correct identity
\item $EER$ or Crossover Error Rate ($CER$): probability that $FMR=FNMR$ or False Accept Rate ($FAR$) = False Reject Rate ($FRR$), where FAR and FRR are system-level errors
\item True Acceptance Rate ($TAR$) which is $1-FRR$; and Weighted Error Rate ($WER$): the weighted sum of $FMR$ and $FNMR$ \cite{Wayman1999}
\end{itemize}

Several curves have also been proposed in order to measure system performance more comprehensively, such as:
\begin{itemize}
\item Receiver Operating Characteristic (ROC) which plots $FNMR$, $FRR$ or $TAR$ on Y-axis and $FMR$ or $FAR$ in X-axis
\item Detection Error Trade-off ($DET$) curve which uses nonlinearly scaled axes to show the regions of error rates of interest
\item Expected Performance Curve ($EPC$) which uses a performance criterion like $FMR$ to measure performance in terms of $FNMR$ or vice-versa
\end{itemize}
The first two curves are a posteriori as the evaluation is dependent on previous knowledge, and the last is a priori since it is independent of prior knowledge. A more comprehensive study was incorporated in \cite{Belen2013}, where TBS performance evaluation was restructured considering the international Common Criteria (CC) for Information Technology (IT) Security Evaluation and its Common Evaluation Methodology (CEM) guidelines, viewing TBS as IT systems.

This paper classifies TBS into \textit{Verification} and \textit{Identification} systems, the latter focused on determining the identity of an individual who may belong to the database (closed-set) or not (open-set).
\textbf{Closed-set system performance evaluation} is done using a \textit{Cumulative Match Characteristic (CMC)} curve proposed in \cite{DeCann2013}, mostly used for systems which generate an ordered list of matches between the test subject and existing samples in the database, sorted from most likely to the least. Each of these matches are labeled as a rank. The first match, which is most likely, is called $Rank$-$1$, followed by the second most likely match labeled as $Rank$-$2$, and so on. CMC plots probability value on the Y-axis against rank on the X-axis. The probability value for a given rank $k$, known as "identification rate for rank $k$", depicts the percentage of time when the system correctly identifies the test individual within the first $k$ ranks. Ideally, the system is expected to identify an individual in the first attempt (first rank). Hence, $Rank$-$1$ CMC is ideal. However, in real-world conditions, systems tend to have identification rates closer to $1$ as the ranks increase. As a measure of performance, the system whose identification rate hits $1$ for the least rank value is deemed to be better. This metric is more widely used for face and iris recognition systems. \textbf{Open-set identification} is done using two approaches: exhaustive comparison where the system compares the test sample with all existing samples in the database, and retrieval-based systems where it employs two subsystems for filtering the samples to allow only those with a "score" above the predefined threshold to be considered for the matching process, and the actual matching with test sample. This class of methods consider two more metrics: Detection and Identification Rate ($DIR$), and False Alarm Rate.

\vspace{-0.3cm}
\subsection{Performance Evaluation Metrics for WBS}\label{sec:3.2}
The performance of WBS under real-world conditions is more complex to measure owing to multiple external factors which vary with the type of modality, such as ambient noise, distractions, poor connectivity between the sensors and sink, lighting conditions, stress, and anxiety. Hence, most research efforts attempt to quantify their performance under laboratory or experimental conditions, making it idealistic.

Nevertheless, performance evaluation of WBS is very nascent. Metrics identified for TBS such as accuracy, FAR, FRR, ERR, and DIR are also applicable for WBS. However, the term "accuracy" has gained a deeper significance among others in wearable context, where it includes not just accuracy in the measurement of the signals, but also the confidence-level, immunity against external and bodily disturbances, quality of operation among sensors, communication latency and throughput, quality of assurance of results, efficient WBAN management to avoid congestion and collision, and energy consumption rate \cite{PwC2015,PwC2014}. In a preliminary survey conducted at the Biometrics Institute Asia Pacific Conference in May, 2016, fifty-four professionals were asked to provide their inputs on the potential applications for WBS, some crucial concerns they saw as hindrances to widespread adoption of the technology, and some potential formats in which they could be made available in the future \cite{Unisys2016}. This paper utilizes the results from the study to propose more metrics with respect to the components identified in the WBS system model earlier.

In literature, much emphasis has been laid on form factor and size as contributing factors towards achieving optimal or near-optimal performance, considering that the performance of WBS decrease with decreasing form factor. While reduced form-factor could sometimes imply greater comfort for the users, it limits local computation capability, necessitating the use of signal processing, feature extraction and matching to be located external to WBAN. This in-turn exposes the WBS to threats that TBS did not have to deal with. Before reviewing the evaluation metrics, the factors on which wearable performance depend are described below in brief.

\begin{enumerate}
\item \textbf{Physiology:} Modalities are sensitive at different levels to different factors such as skin complexion, body shape and size. For example, the level of fat under the skin could affect the measurement of EMG signals, but it might not affect the measurement of a signal like EEG. Similarly, skin complexion could change the level of absorbed light (used by PPG)
\item \textbf{Number and placement of sensors} directly correlates with the quality of signal measurement. Improper placement or implantation of sensors leads to weak or erroneous measurement, affecting quality and accuracy, and hence performance
\item \textbf{Changes induced by mobility:} Quality of operation must not vary beyond an acceptable range when users are subject to movement, including rigorous physical exercise, true especially for ECG, PPG and EEG that are prone to interference with external noise
\item \textbf{Environmental elements:} External noise due to ambient elements could significantly corrupt the measured signals. A technique called Active Signal Characterization (ASC), proposed by Valencell, measures biological, motion, and environmental signals as they come through optical and accelerometer sensors \cite{Leb2016}. The required signal is then filtered to remove noise related to motion and other external signals. The separated noise could be used as supplementary data to deliver crucial insights into the user environment and motion-related features. A preliminary study has also been conducted on PPG \cite{Val2016}
\item \textbf{Crossovers:} Periodic movements made by a body could be mistaken by the system to represent a biometric modality. For example, the step-rate measured during activities like jogging or running could be mistaken for heart-rate and lead to miscalculations or wrongful verification/identification, both which bring down performance
\item \textbf{Performance of the matching classifiers:} Whether one-class or multi-class, feature matching algorithms have their pros and cons with respect to performance. No matter how good a signal measurement, a less accurate classifier with poorly fit model could yield substandard results. The classifier model's performance for skewed training data (classification bias), precision, recall, and F1-score are some additional performance metrics that could be used to indirectly measure WBS performance \cite{Diaz2016}
\item \textbf{Handling heterogeneity:} There are multiple types of wearable sensors that comprise WBAN, but measurement, signal preprocessing, feature extraction, selection and matching techniques for each of them vary significantly. 
Hence, the features that extractor and selector models look for also vary. However, in a comprehensive wearable environment, the extractor/selector and matcher are all embedded within a single device. Hence, the models used for performing such tasks must be adaptive to more than one type of signal, and the changing tolerance levels to different signals affects performance
\end{enumerate}

Based on these factors, many performance evaluation metrics have been proposed recently in literature, including some patents \cite{Eric2015,Steven2014,Arxan2015,Awais2016}. They have all been investigated for healthcare and personal fitness, specifically for PPG. It is noteworthy that, like characteristics, metrics for TBS performance evaluation are also applicable for WBS. Following are the metrics unique to WBS.
\begin{itemize}
\item \textbf{Accuracy:} Affected by factors 1-7; its calculation could be subjective due to the involvement of many subject-variant factors. Companies such as Valencell have attempted to deliver Precision Wearable Biometrics that account for these factors. Much emphasis has been laid on wearables for healthcare and fitness applications, but not user security
\item \textbf{Flexibility:} Affected by factors 1, 2, 7; considering the dynamic and highly sensitive nature of WBS, much performance analysis has been typically done through validation testing by classes of users representing different physiologies such as skin tone, complexion and texture, body shape and size, and much more \cite{Kos2017}. It has become a common practice employed by most wearable companies prior to releasing products. In other words, the WBS should be flexible enough to be used by majority of the human population
\item \textbf{Interoperability:} Affected by factors 3, 4, 7; WBS must actively communicate either with other wearables within the same WBAN or to the sink(s). They must not only measure data with good quality, but also transmit them to the sink without losses or corruption. WBAN communication strategies discussed earlier in Section \ref{sec:2.1} provide an insight into the interoperability of WBS
\item \textbf{Security:} Affected by factors 2, 4, 5; besides lossless and reliable communication, security is also key to performance. Encryption might not be suitable to secure signals from wearable sensors as it demands additional computation power. Efficient resource-aware key management and signal data scrambling methods are two alternatives, where descrambling could be performed by the sink during signal preprocessing
\item \textbf{Resource allocation:} Affected by 2, 4, 5; prolonged battery life is one of the highly anticipated deliverables for WBS. Hence, the extent to which WBAN nodes manage their computation and communication powers to maximize functioning while minimizing consumption could be a metric for evaluating performance
\item \textbf{Versatility:} Affected by factors 1-3, 7; it is the extent to which a WBS can be effectively molded into different consumer-friendly forms that encourage longer battery life and reduced form factor. People are fashion-conscious, and hence, WBS must not only be non-invasive and comfortable, but also be elegantly designed with appealing aesthetics in such a way that performance is not hindered
\item \textbf{Power Consumption:} Affected by factors 2, 4-7; in order to optimize power consumption, the amount of energy expended by sensors for every bit of information could determine their performance \cite{Wendi2000}. Furthermore, as shown by \cite{Filippo2013} for Bluetooth-WBAN, synchronization strategies could help improve accuracy and limit unnecessary power consumption, thereby improving overall performance. Wake-up Radios (WURs) were proposed to listen to wireless channels in a power-efficient way through preamble sampling and continuous channel listening \cite{Stevan2012}
\item \textbf{Network Efficiency:} Affected by factors 2, 4, 5; optimal use of communication bandwidth is a key concern for nodes within WBAN, between WBAN nodes and sink or, in some cases, between sink and cloud. Since most wearables today use wireless communication like WiFi, Bluetooth and cellular networks, they operate in the same frequency bands as other devices like mobile phones, laptops, and smart home appliances. While temporary solutions such as freeing up more bandwidth, spectrum sharing, and dynamic allocation and deallocation of bandwidth depending on the usage have been implemented, underlying bottlenecks of network congestion, noise, and subsequently performance degradation persist. Visible spectrum was proposed to establish wireless information transfer via Light Emitting Diodes (LEDs) \cite{Bhalerao2013}. Multi-tiered network architecture was also proposed, bolstering the emerging fifth generation (5G) mobile-communication systems where there exists one device, known as "seed", that connects to the internet and relays common information to the nodes subscribed to it \cite{Hossain2014}. Dynamic use of bandwidth can be feasible through the use of Cognitive Radios, which hop into underutilized bandwidths, lowering latency and maximizing performance \cite{Haykin2005}
\item \textbf{Spectral Efficiency:} Affected by factors 2, 4, 5; measured in bits/s/Hz, it refers to the extent to which a physical or MAC layer protocol can effectively use the limited frequency spectrum bandwidth available (Hz) to maximize its information rate (bits/s). For WBS, this is very important given their limited resources and bandwidth. Spectral efficiency $\eta$ has a probable mathematical formulation as shown below, where $N_R$ and $N_T$ are the number of receivers (sink) and transmitters (sensors) respectively; $G_{ij}$ is the goodput (transmission of useful information bits between transmitter $i$ and receiver $j$ per unit time); $D_{ij}$ is distance between transmitter $i$ and receiver $j$; $U_{ij}$ is mean societal value received by transmitter $i$ from receiver $j$ in return for every bit transmitted, where the societal value includes economic, social and environmental benefits; and $A$ and $B$ are area and bandwidth, respectively, in which transmitter $i$ and receiver $j$ operate \cite{Stanciu2012}:
\begin{equation}
\label{eqn:01}
\eta=\sum_{j=1}^{N_R}\sum_{i=1}^{N_T}\frac{G_{ij}D_{ij}U_{ij}}{AB}
\end{equation}
\end{itemize}

\textbf{Key Takeaway Points:} The following are the key takeaway points from this section:
\begin{enumerate}
\item TBS performance has been conventionally measured using metrics like FAR, FRR, FMR, FNMR, EER/CER, TAR, ROC, DET, EPC
\item Although the metrics used to evaluate TBS can be extended to evaluate WBS, the latter have more metrics owing to their differing operational dynamics
\item Some of the key metrics that can be used to evaluate the performance of WBS include system accuracy, operation flexibility, device and application-level interoperability, systemic security, functional versatility, resource-efficiency, energy-efficiency, network efficiency, and spectral efficiency
\item WBS performance measured by each of the above metrics is in-turn impacted by one or more of various factors such as physiology, placement and orientation of sensors, external environment, crossovers, system component performance, and heterogeneity in sensing and computation
\end{enumerate}

\vspace{-0.2cm}
\section{Threats, Attacks and Defenses for TBS and WBS}\label{sec:5}
Before moving ahead, a primer on the terms related to biometric security is briefly presented. A \textit{Threat} is anything that has the potential to inflict serious harm to the technology of concern. A \textit{vulnerability} is a weakness that when successfully exploited manifests the threat into an attack \cite{Abdul2009}. An \textit{attack vector} is a means by which a malicious entity can compromise a system by exploiting its vulnerabilities for a malicious outcome \cite{Schatten2009}. An \textit{attack} is the execution of an attack vector by an adversary who successfully exploits a series of vulnerabilities in the concerned domain. A coordinated attack comprises multiple attacks, sequential or parallel, represented using attack vectors that may exploit the same vulnerabilities. An attack is applicable to both TBS as well as WBS in two ways: system attacks that tamper the hardware or firmware, or pattern recognition attacks where the feature extraction and feature matching modules are harmed. While system attacks are applicable to any domain, pattern recognition attacks target feature selection and extraction. It can be said that TBS and WBS are both prone to system attacks, but it is harder to conduct pattern recognition attacks on WBS. However, it is not impossible to do so, as will be discussed in Section \ref{sec:4}.

The threats, attacks, and defense landscapes can be categorized for the purposes of this survey into different classes. The first, called \textit{Technology Attack ($T$)}, considers attacks that exploit vulnerabilities of the system's technology, while the second, called \textit{Modality Attack ($M$)}, considers attacks that exploit vulnerabilities of biometric modalities used by the technology. When an adversary exploits vulnerabilities of both technology as well as modalities, a \textit{Hybrid Attack ($H$)} could be realized. These attacks can be applied to both TBS as well as WBS.
There are also different \textit{Agents} which enforce a threat into an attack: impostor that deliberately or accidentally pretends to be the authorized entity, attacker that intends to access or compromise the technology with malicious intent, snooper that intends to access or compromise the technology with no malice, and erroneous that compromises the technology accidentally. Sometimes, these agents could also be non-human \cite{Abdul2009}. Penetration testing is conducted to discover hidden vulnerabilities and establish attack vectors that can then be mitigated \cite{biggio2015}.

\vspace{-0.2cm}
\subsection{Threats, Attacks and Defenses for TBS}\label{sec:4.1}
\begin{figure}
  \includegraphics[scale=0.3]{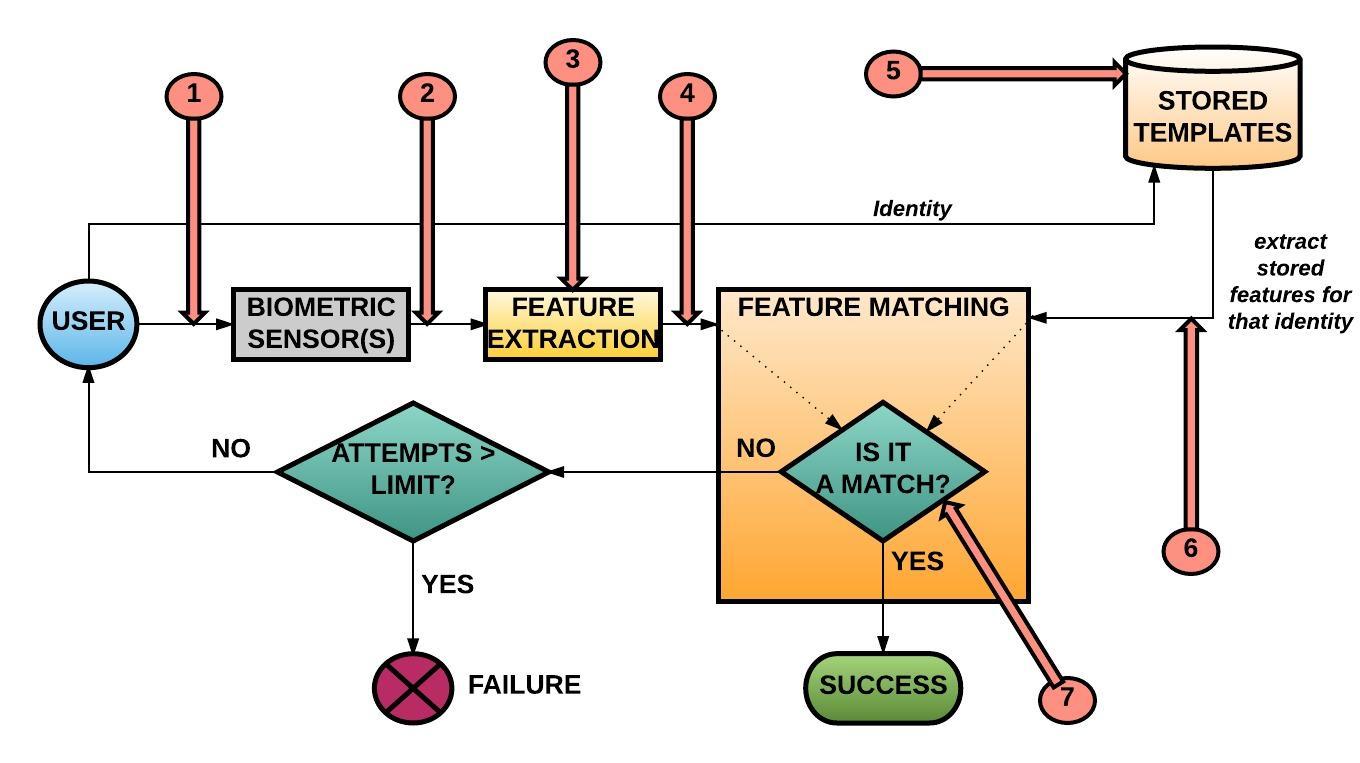}
  \caption{Block diagram of attack model for TBS.}
  \label{fig:five}
\end{figure}

Figure \ref{fig:five} shows the system model of TBS with different points of attacks, as identified by the different works in literature \cite{Delehante2011,Parvez2014,Dellys2013,Adler2008}. In this figure, only the authentication phase of the system is shown, since enrollment is subsumed in the authentication except for the template creation stage, the security-related significance of which was already described in Section \ref{sec:1}. Considering how various components interact with each other, a successful attack on one component could pave way for subsequent compromise of other components. It is to be understood that the categorization of attacks described earlier in this section is made with a certain degree of independence between them. For example, a successful Modality attack does not guarantee a successful compromise of the entire system, since the technology might still remain unaffected. Similarly, a Technology attack might be conducted without directly compromising any of the modalities. Those attacks which require to disrupt both technology as well as modality are categorized as Hybrid to avoid any confusion. While a successful Modality attack affects confidentiality but not necessarily integrity and availability, a successful Technology attack impacts integrity and availability but not necessarily confidentiality. A Hybrid attack compromises all three cornerstones of security.

Table \ref{tab:three} summarizes the attacks, corresponding threats and defenses against them. Analysis of system-specific attacks and defenses have been extensively studied in literature \cite{Denis2006,MdAbed2012}.

\begin{longtable}{|>{\raggedright}p{2cm}|>{\raggedright}p{3cm}|>{\raggedright}p{4cm}|>{\raggedright\arraybackslash}p{3.5cm}|}
\caption{Summary of class-wise TBS threats, attacks and defense mechanisms}\\\hline
\multicolumn{1}{c}{\textbf{Class}} & \multicolumn{1}{c}{\textbf{Attacks}} & \multicolumn{1}{c}{\textbf{Threats, Vulnerabilities}} & \multicolumn{1}{c}{\textbf{Defenses}}\\\hline
\endfirsthead
\multicolumn{4}{@{}l}{\ldots continued}\\\hline
\multicolumn{1}{c}{\textbf{Class}} & \multicolumn{1}{c}{\textbf{Attacks}} & \multicolumn{1}{c}{\textbf{Threats, Vulnerabilities}} & \multicolumn{1}{c}{\textbf{Defenses}}\\\hline
\endhead
\hline
\multicolumn{4}{r@{}}{continued \ldots}\    \endfoot
\hline
\endlastfoot
  M \linebreak \cite{Jain2003,Erdogmus2014,Chak2014,ISOIECPA,Galbally2007,Galbally2012,Galbally2014c,Rahman2015,PADfinger1,PADfinger2,PADfinger3,ISOIECPAD,ISOIECPADSt,Zamboni2013}
& Presentation attack, \linebreak Collusion, \linebreak Coercion & Easily reproducible, \linebreak Poor manual supervision, \linebreak Poor fault tolerance, \linebreak Gullible to close relatives, \linebreak Residuals
& Supervised enrollment, \linebreak PAD, \linebreak Access controls\\\hline
  Sys \linebreak \cite{ICITIS2007,Paillier,Lim2013,Reddy2016,Trabelsi2014,Kaur2010,AKJain2005,Dorn2012,Mehta2016,Khalaf2015,Abhyankar2010,Tuyls2005,Chang2004,Uludag2004,Anil2008}
& Physical tamper, \linebreak Replay, \linebreak Man-in-the-Middle, \linebreak Bad data injection, \linebreak Template substitute, \linebreak Template compromise, \linebreak Denial of Service
& Lack of policy enforcements, \linebreak Sensor memory/ circuit limits, \linebreak Flaw in sensor software, \linebreak Poor key management, \linebreak Weak/no encryption, \linebreak Vulnerable communication, \linebreak No perfect forward secrecy, \linebreak Lack of interoperability between other systems, Susceptibility to Trojan horse, \linebreak Trade-off between template security and system performance, \linebreak Storage channel interception, \linebreak Key discovery
& Firewalls, \linebreak Physical security, \linebreak Fall-back systems for fault tolerance, \linebreak Use of Session keys and timestamps, \linebreak Digital signatures, \linebreak Random/ cued challenges, \linebreak Multi-factor authentication, \linebreak Biometric cryptosystems, \linebreak Template encryption, \linebreak Template fusion, \linebreak One-time biometrics\\\hline
  PR \linebreak \cite{Lafikh2015,Juels2002,Zhou2009,Adler2009,Lim2016,Rathgeb2011,Nagar2010,Teoh2015,Yang2007,Makinde2014,Surmacz2013,Ballard2008,Ohaeri2005,Shanthini2012,Akhtar2012,Cui2008}
& Characterize feature extractor, \linebreak Feature replay, \linebreak Feature correlation, \linebreak Trojan horse to alter match scores
& Known template storage format, \linebreak Information of data type created by feature extractor, \linebreak Correlation among features, \linebreak Matcher and decision-maker program error, \linebreak Poor exception handling, \linebreak Missing match upper bounds
& Use biometric feature entropy, \linebreak Revocable biometrics, \linebreak Feature randomization from multiple input sets, \linebreak Effective debugging with exhaustive use-case testing, \linebreak Match-score quantization\\\hline
  H \linebreak \cite{Alaswad2014,Maiorana2015,Galbally2010,Alex2016,Diaz2016,Nelder1965,He2010,Hanmandlu2011,Yangsen2012}
& Buffer Overflow, \linebreak Enrolling crafted samples to matcher, \linebreak Hill climbing
& Fraud during enrollment
& Feature fusion from multiple modalities
\label{tab:two}
\end{longtable}

\subsubsection{$M$ Threats, Attacks and Defenses.}\label{sec:5.1.1} This class refers to attacks that exploit the vulnerabilities of the users or their modalities to successfully penetrate into the system. Spoofing or identity (ID) theft is the oldest form of attack (also called a Direct or Presentation Attack) where modalities such as fingerprints, iris, signature and others can be recreated or spoofed using gummy fingers, high resolution color printouts, 3D robotic eyes, and much more \cite{Jain2003,Erdogmus2014,Chak2014,ISOIECPA}. In general, to protect against Class 1 attacks, proper enforcement of enterprise-level security such as access controls must be enabled. Additionally, security policies must be strictly enforced to avoid creating loopholes. Physical hardware of TBS must be capable of working without significant loss of performance in the event of an attack or outage \cite{Zamboni2013}. Targeted direct attacks have lately been minimized by the use of sophisticated sensors equipped with Presentation Attack Detection - PAD (also called live-ness or vitality detection, or anti-spoofing) techniques \cite{Galbally2007,Galbally2012,Galbally2014c,Rahman2015,PADfinger1,PADfinger2,PADfinger3,ISOIECPAD,ISOIECPADSt}.

\subsubsection{$Sys$ Threats, Attacks and Defenses.}\label{sec:5.1.2} This class of attacks target the system components by exploiting vulnerabilities at different points identified in the system model (Figure \ref{fig:one}) except the feature extractor and matcher: communication channels, modality residuals, and the template storage unit. Two forms of eavesdropping attacks exist. Passive eavesdropping intercepts the vulnerable communication channel between sensor and feature extractor modules but does not alter or steal it; active eavesdropping like Man-in-the-Middle (MITM) and storage channel interception usurp, swap, or corrupt legitimate data to disrupt rightful operation of the system \cite{ICITIS2007}. While only a few employ encryption when sending the captured biometric signals, they do not enforce forward secrecy (a method to ensure the non-compromise of previously secure events even if the current event is compromised) \cite{Paillier}. Attacks and defenses related to biometric templates discussed in Section \ref{sec:1} are applicable here. Since TBS are database-oriented, they need to query the stored template in order to make the comparison, which an attacker could modify. If the encryption is not strong, the attacker can replace the query with malware to corrupt the database \cite{Lim2013}. Template swapping is a special case of substitution. It deals with replacing a legitimate template with that of any user, even external to the system \cite{Kaur2010,AKJain2005}. Alternatively, template could be compromised by exploiting weaknesses in the database architecture or schema, enabling the attacker to insert, update or delete templates of legitimate users. DoS can be executed by feeding TBS with an overwhelming number of feature samples through Bad Data Injection (BDI), MITM, or replay \cite{Trabelsi2014}.

Specific defense mechanisms such as tagging input signals with timestamps and using session keys to ensure forward secrecy have been developed \cite{Reddy2016}. Template encryption, where secure sketch schemes, homomorphic encryption, and chaotic theory, could be used, considering standard hashing and encryption methods which are applicable for other types of data fail for biometric templates \cite{Dorn2012,Mehta2016}. A modified Hill Cipher algorithm for encryption and a combinatorial Discrete Cosine and Discrete Wavelet Transforms for concealment were proposed that led to an overall improvement in template storage \cite{Khalaf2015}. Templates could be alternatively stored separately on smart cards, backed by biometric cryptosystems. However, this imposes an additional restriction of users having to carry cards all the time and increases the overall system complexity. One-time biometrics deter storage channel interception by using statistical learning to create biometric representations, and then applies chaotic mixing to generate an encrypted template, constituting a self-generated, dynamic helper data \cite{Abhyankar2010}. Encrypted template is then decrypted into constituent biometric representations using well-trained Hidden Markov Models (HMMs) and iterative Blind Source Separation before being fed into fuzzy matcher. A multi-factor authentication can also provide defense against MITM attacks by combining modality samples with PINs or passwords \cite{Anil2008}. However, such additional authentication mechanisms could prove counterproductive to the reason biometrics were introduced. Biometric cryptosystems, which generate a helper data using a secret key and biometric features, may also be used to prevent MITM attacks \cite{Uludag2004}. Here, only the helper data, which by itself is useless to an attacker, is stored by the system, while the secret key is reconstructed during authentication using extracted features and helper data \cite{Tuyls2005,Chang2004}.

\subsubsection{$PR$ Threats, Attacks and Defenses.}\label{sec:5.1.3} Attacks on feature extractor and matcher can be regarded mostly as $PR$ in nature. Since randomness in modality features is hard to achieve for cryptosystems, fuzzy-based extractors using shielding functions, fuzzy commitment, and fuzzy vault schemes were proposed and applied to the key binding process \cite{Lafikh2015,Juels2002}. However, attacks using feature correlation are shown to significantly reduce performance \cite{Zhou2009}. If the attacker has prior knowledge about the feature extractor, optimization methods to estimate the unknowns can be devised using the knowns, exploiting statistical dependencies. Further, knowledge of feature extracting algorithms such as PCA and LDA which use the entire biometric sample to construct feature vectors, and Gabor filters and HMMs which select specific features of modalities prior to forming the feature vector, can be used to exploit the correlation between features of modalities that the extractors also use \cite{Adler2009}. False Data Injection (FDI) could corrupt extracted features by the addition of random noise or intelligent data. Such attacks could either be aimed at feeding a large number of erroneous feature vectors to the matcher that would increase its FRR beyond acceptable limits, or at manipulating the vector to circumvent or bypass the matcher.

Probable defense mechanisms include measuring feature entropy to gauge the level of uniqueness of the modality and also the strength of cryptosystems for guaranteeing privacy \cite{Lim2016}. Cancelable or Revocable biometrics, also referred to as template transformation, distort input features by a specific function by applying Gaussian noise models, and use multiple distorted features for different levels of authentication \cite{Rathgeb2011,Nagar2010}. They store not the original features but only the distorted ones. However, they have a potential of increasing the system's FRR considering the inherent feature variability among modalities. Feature extractors take this into account using error correction codes, adjustable filters, correlation, or quantization \cite{Teoh2015,Yang2007,Makinde2014,Surmacz2013}. Alternatively, randomization of biometrics could be used, where multiple samples of a biometric signal is taken (the number of samples varies), and a cumulative average of the samples is used to minimize the intra-class variance \cite{Ballard2008}. Multimodal feature fusion is also proposed, which differs from revocable biometrics by applying transformations and distortions to a single modality feature set to produce multiple feature vectors. Feature fusion has been successfully applied to feature extractors and tested for its robustness against various attacks including spoofing and replay of extracted features, and potentially feature correlation attacks \cite{Ohaeri2005,Shanthini2012,Akhtar2012,Cui2008}. However, practical significance of multi-factor authentication, revocable biometrics and biometric cryptosystems is limited.

\subsubsection{$H$ Threats, Attacks and Defenses.}\label{sec:5.1.4} Attacks that are a hybrid of $M$ and $Sys$: buffer overflow and residuals. Buffer overflow involves a system interacting with an external environment seeking inputs, in this case, the sensor(s) \cite{Alaswad2014}. Flaws in sensor memory allocation protocols could be exploited by such an attack to overwrite codes important for system's rightful operation. System authentication could be bypassed in some cases. Further, residuals (latent fingerprints, signature imprints, or pressure points on specific keys in keystroke recognition systems) can be exploited to conduct replay attacks \cite{Alex2016}. A Hill Climbing (HC) attack is a hybrid of $M$, $Sys$ and $PR$ attacks, made more powerful by Nelder-Mead simplex optimization \cite{Nelder1965}. It creates an application that sends random templates to the system, disturbed iteratively \cite{Diaz2016,Galbally2010}. The system works by reading output match score and proceeding with perturbed template only after score surpasses the acceptance threshold. Successful applications of HCA to TBS exist in literature \cite{Maiorana2015,Galbally2010}.

Since HCA targets matching scores of feature matching algorithms, defense mechanisms aim to immunize the matcher. Score-level fusion techniques namely, SVMs, likelihood ratio-based fusion, and sum rule-based method preceded by normalization, were analyzed for their performance and accuracy for multimodal biometric systems comprising fingerprints, face and finger vein \cite{He2010}. Another contemporary method proposed the use of triangular norms to make score-level fusion faster and computationally efficient, again for multimodal systems \cite{Hanmandlu2011}. HCA was applied for online signature TBS by modifying initialization, restart and centroid computation steps of traditional Nelder-Mead algorithm. It also proposed a Llyod-Max non-uniform score quantizer to determine quantization levels such that the Mean Square Error (MSE) between original and quantized versions is minimized. Additionally, decision-level template fusion was shown to degrade performance least when compared to fusion at sample, instance, or algorithmic levels \cite{Yangsen2012}. Since matching and decision-making modules of the system contain program codes that have access to crucial parameters besides the match score, like FRR, FAR, FNMR, and EER, ineffective program blocks could be compromised through the use of Trojan horse.

\begin{figure}[!b]
  \includegraphics[scale=0.3]{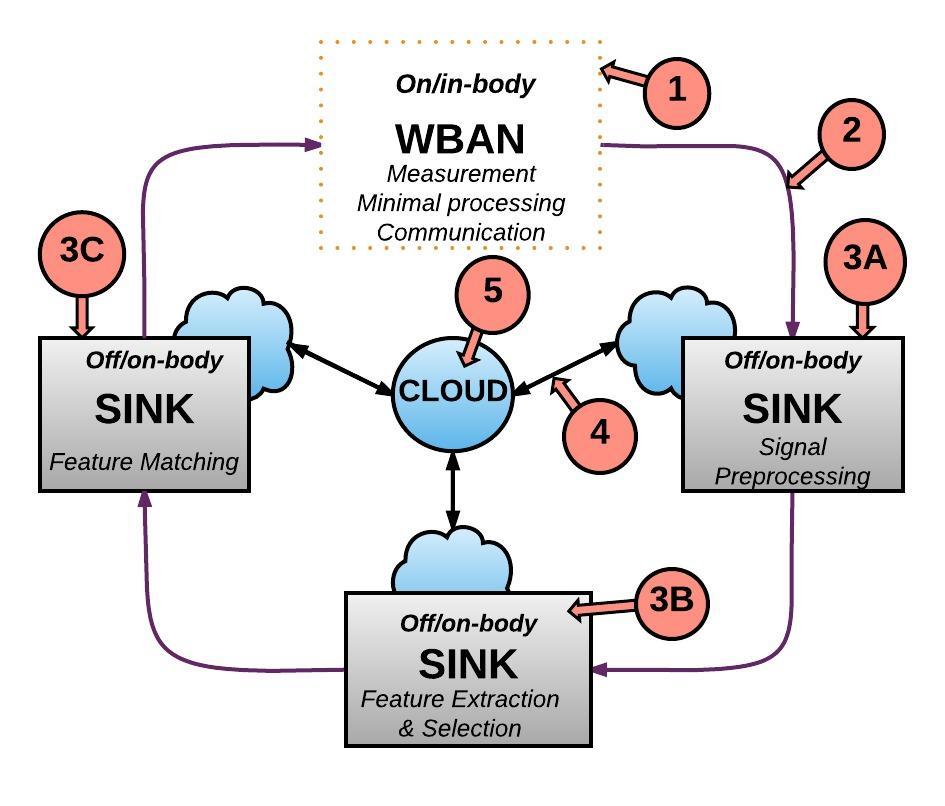}
  \caption{Block diagram of attack model for WBS.}
  \label{fig:six}
\end{figure}

\subsection{Threats, Attacks and Defenses for WBS} \label{sec:4.2}
WBS are associated with a new dimension of threats and vulnerabilities. Figure \ref{fig:six} shows their generic attack model. In WBS domain, Class $M$ attacks exploit vulnerabilities of modalities and users, while $Sys$ attacks are surveyed under two classes: $Sys$-$WBAN$ and $Sys$-$Sink$. Attacks on WBAN sensors, intra-body communication, and the communication channel between WBAN and sink come under $Sys$-$WBAN$ while those that target the sink's signal preprocessing unit, communication between sink and cloud, and cloud storage infrastructure itself. Class $PR$ attacks target the sink's feature extraction and selection, and feature matching. Table \ref{tab:three} summarizes threats and vulnerabilities, attacks and corresponding defenses for WBS, organized class-wise.

\begin{longtable}{|>{\raggedright}p{1.5cm}|>{\raggedright}p{3cm}|>{\raggedright}p{4cm}|>{\raggedright\arraybackslash}p{4cm}|}
\caption{Summary of class-wise WBS threats, attacks and defense mechanisms}\\\hline
\multicolumn{1}{c}{\textbf{Class}} & \multicolumn{1}{c}{\textbf{Attacks}} & \multicolumn{1}{c}{\textbf{Threats, Vulnerabilities}} & \multicolumn{1}{c}{\textbf{Defenses}}\\\hline
\endfirsthead
\multicolumn{4}{@{}l}{\ldots continued}\\\hline
\multicolumn{1}{c}{\textbf{Class}} & \multicolumn{1}{c}{\textbf{Attacks}} & \multicolumn{1}{c}{\textbf{Threats, Vulnerabilities}} & \multicolumn{1}{c}{\textbf{Defenses}}\\\hline
\endhead
\hline
\multicolumn{4}{r@{}}{continued \ldots}\    \endfoot
\hline
\endlastfoot
  M \cite{Cai2012,Beltramelli2015}
& Motion-based keystroke inference, \linebreak Collusion, \linebreak Coercion & Poor manual supervision, \linebreak Poor fault tolerance, \linebreak Gullible to close relatives, \linebreak Residuals
& Supervised enrollment, \linebreak PAD, \linebreak Access controls\\\hline
  Sys-WBAN \cite{Wang2016,Karlof2003,Bysani2011,Wang2016,Newsome2004,Ali2013,Bui2008,Mana2011,Zhu2011,Xu2011,Seyedi2015,Nie2015}
& Direct attacks, \linebreak MITM, \linebreak (D)DoS, \linebreak Counterfeit sensors, \linebreak Selective forwarding, \linebreak Jamming, \linebreak Brute-force search, \linebreak Sybil/Wormhole
& Sensors transmit only useful information, \linebreak Difficult policy-making, \linebreak Expensive to encrypt, \linebreak Feature entropy, \linebreak WBAN broadcasting, \linebreak Noisy network, \linebreak Half-duplex intra-WBAN communication
& Multi-point fuzzy commitment, \linebreak Cluster-based security, \linebreak MBStar WBAN topology, \linebreak Trust-based key management, \linebreak Elliptic Curve, Cryptography \linebreak Additional hardware filters, \linebreak Defensive jamming, \linebreak Hyper-quiet networks, \linebreak Tamper-proof sensors, \linebreak BCC\\\hline
  Sys-Sink \cite{Austen2015,Vidalis2014,silee2014,silee2013,Ibrahim2016,Tintel2015,Wilkins2014,Someren2015,Weii2014,Duan2005,Ching2016,Naveed2014,Rahmanm2016,Kharpal2015,Ani2016,Zeller2008,Maiti2015,Kocher1999,MArtin2004,Rahmanc2014,Manivannan2016,Linn2011,Manjusha2015,Leeh2015}
& Sniffing, \linebreak Data misuse by third party, \linebreak Threats due to third party, \linebreak Providing false trust credentials, \linebreak SQL injection, \linebreak CSRF, \linebreak Traffic redirection, \linebreak Malware, \linebreak Privacy exposure, \linebreak Session hijack, \linebreak BDI, \linebreak JTAG (-R), \linebreak External device mis-bonding, \linebreak Side-channel attack
& Weak/no WBAN-to-sink authentication, \linebreak Varied off-body sink network dynamics, \linebreak Third party encryption, \linebreak More data logged than revealed, \linebreak Application of similar security levels for different kinds of data, \linebreak Sink broadcasts to cloud, \linebreak TLS could be bypassed, \linebreak Location-related information storage, \linebreak Obtrusive authentication, \linebreak Untrusted applications installed, \linebreak Vulnerabilities of sink, \linebreak Weak device-application bonding policies
& Cluster-based security, \linebreak Proximity detection, \linebreak Mutual authentication, \linebreak Random challenge/ response, \linebreak Network segmentation, \linebreak Administration of backups, \linebreak Dedicated onsite cloud, \linebreak Careful policy inspection, \linebreak Data protection, \linebreak Context-based data security, \linebreak Trust revocation, \linebreak HTTPS with SSL, \linebreak Network segmentation, \linebreak Data compartmentalization, \linebreak Symmetric encryption, \linebreak Timestamps to obfuscate patterns, \linebreak OS-level device-application bonding \\\hline
  PR \cite{Nisha2014,Kelsey1998,JChuang2014}
& Cryptanalytic attacks
& Noise/ redundancy in signals, \linebreak Collection of geolocation and time-synchronized information, \linebreak Time-variance of signals
& Matching-level fusion, \linebreak Multi-factor authentication, \linebreak Feature-level fusion, \linebreak Attribute-based encryption, \linebreak Case-based reasoning, \linebreak Spectral analysis of signals
\label{tab:three}
\end{longtable}

\subsubsection{$M$ Threats, Attacks and Defenses.}\label{sec:5.2.1} At the modality-level, WBS exhibit slight variations in attack surface than TBS. While direct attacks are possible in WBS too, they are not conducted using shoulder-surfing or spoofing of templates, since WBS employ passive authentication. Hence, such attacks are conducted at the system level (Class $Sys$-$WBAN$). \textbf{Collusion} and \textbf{coercion}, however, are still prevalent techniques that employ social engineering skills to deceive a legitimate WBS into authenticating or identifying an attacker. A \textbf{motion-based keystroke inference attack} on smartphones was demonstrated by \cite{Cai2012}, where its feasibility was shown despite the influence of noise, device dimensions and screen orientation. Using Long Short-Term Memory (LSTM) deep neural networks, motion sensors of wearable wristbands like smartwatches could be used to infer keys typed by the owner on other devices like PC or cellphone keypad for spying \cite{Beltramelli2015}.

\subsubsection{$Sys$-$WBAN$ Threats, Attacks and Defenses.}\label{sec:5.2.2} This class of attacks target WBAN and the communication between WBAN and sink. Owing to stringent energy and resource limits, WBAN sensors transmit only useful information, and go through sleep and wakeup cycles to conserve energy, relaying messages through half-duplex wireless broadcasts. By virtue of this, \textbf{direct attacks} by eavesdropping are prevalent. Thorough network reconnaissance could make an external wireless node privy to signals being exchanged between sensors, and sometimes even between sensors and sink. Decoded signals could reveal sensitive information that could then be stolen by the adversary, or even modified to make sensors act anomalously, causing life-threatening behaviors. Sometimes, counterfeited sensors could be deployed into WBAN which are prone to \textbf{jamming} attack where a malicious node reduces availability by causing collision to each packet of interest \cite{Wang2016}, \textbf{Sybil attack} where a sensor node exhibits dynamic personality by falsely claiming the identities of other surrounding nodes within the same WBAN through impersonation \cite{Newsome2004}, or \textbf{wormhole attack} where a sensor node could trick other nodes in the network into thinking it is only a few hops away from them when in reality it could be otherwise, thereby not only confusing the routing algorithms, but also causing more energy and resource consumption \cite{Karlof2003}. Counterfeited or compromised sensors could also be manipulated to conduct \textbf{selective forwarding attacks}, where the node intercepts only specific packets of data but not all, thus prompting re-transfer or restart of packets that in-turn drains energy \cite{Bysani2011}.
Most WBS use Bluetooth, WiFi, and ZigBee for communicating with their sinks \cite{Austen2015}. For higher performance and comfort, most commercial WBS transfer sensitive geolocation in clear text, which could be easily intercepted through \textbf{brute-force} attacks. Most WBAN-sink communications only employ sink-to-WBAN authentication to minimize the overhead on WBAN. However, this makes the entire communication link vulnerable to \textbf{MITM} and \textbf{salami thefts}, the latter of which steals small chunks of sensitive data unrecognizable individually but compromises over time. No work in literature has studied the effect of salami attack on WBS, but an assessment of salami attacks and ID thefts on IoT, a superset of WBS, was conducted by \cite{Vidalis2014}.

A \textbf{multi-point Fuzzy commitment} scheme for key management using ECG was proposed, and its performance evaluated by augmenting Gray coding into its binary encoder and error correcting codes to fine-tune accuracy \cite{Bui2008}. A unique \textbf{data scrambling} approach using interpolation and random sampling was also applied instead of conventional symmetric cryptographic techniques. However, the practicality of this strategy in the presence of fading and distortions was only briefly addressed. Another energy-efficient key management and refresh scheme using \textbf{multiple clusters} was proposed, which used both predetermined as well randomly generated ECG keys for creating hybrid security \cite{Ali2013}. This technique is applicable better to WBANs than data scrambling. A more reliable and secure protocol for the inherent star topology of WBANs was proposed, termed as \textbf{MBStar} \cite{Zhu2011}, addressing the problem of long hyper-period communications required by TDMA in the MAC layer under varying schedule profiles by keeping a global hyper-period schedule on the gateway/sink side and assign node-specific local schedules with conflict resolution. It demonstrated co-existence functionalities with other standard protocols like Bluetooth, ZigBee and WiFi. A \textbf{trust key management} scheme for WBAN implemented using ECG utilizes ECG to generate symmetric session keys and manage them for end-to-end communication, also between sensors and sink \cite{Mana2011}. Some approaches advocate the separation of signal measurement and authentication in sensors, where a separate middle-entity called \textbf{Guardian} was proposed for the Implantable Medical Devices (IMDs) scenario to protect against jamming, direct attacks, selective forwarding, and brute-force search through techniques like \textbf{defensive jamming} (jamming sensors when communication link is jammed by an attacker) \cite{Xu2011}. Some defense methods consider non-Radio Frequency (RF) wireless data communication using \textbf{Body Channel Communications} (BCCs), where the human body is converted into a channel/medium to send messages, consuming power less than $1mW$ at rates greater than $100kbps$ \cite{Seyedi2015,Nie2015}.
\textbf{Hyper-quiet network} principles and network segmentation employ layered architecture for isolating network traffic based on the type of data to be transmitted, thereby lowering network congestion. Although encryption is the most sought solution in networks, it is an expensive feature for WBAN-sink communications, especially with WBAN. Traditional Public Key Infrastructure (PKI) use RSA, but other methods such as \textbf{Elliptic Curve Cryptography} (ECC) in conjunction with symmetric encryption have been proposed \cite{silee2014}. In some cases, proximity detection can be used where the distance between WBAN and sink can ascertain whether the signal is fraudulent. However, simple proximity checking might not work in cases where wearables are designed to perform remote services. Hence, \textbf{context-aware proximity checking} is needed. \textbf{Mutual authentication} could also be used where wearable sensors and sink establish handshake to ensure secure communication channels using Physical Unclonable Function (PUF) that employs challenge-response methods to parry replay attacks \cite{silee2013}, and an energy-aware mutual authentication by the use of hash and XOR operations over a star two-tiered topology \cite{Ibrahim2016}.

\vspace{-0.3cm}
\subsubsection{$Sys$-$Sink$ Threats, Attacks, and Defenses.}\label{sec:5.2.3} This class targets the sink's signal processing and computation units, cloud, and communication link between sink and cloud. While some WBS feature local storage and processing capabilities, still many others like Google Glass and Fitbit opt for cloud-based data storage and processing. Although loss of privacy and \textbf{network impersonation} attacks could be conducted when sink communicates with cloud, \textbf{traffic redirection} and \textbf{Cross-Site Request Forgery} (CSRF) attacks are also emergent \cite{Zeller2008}. A greater number of organizations adopt Bring Your Own Device (BYOD) \textbf{policies}, where multiple sinks can access similar services offered by the cloud \cite{Kharpal2015}. Multiple applications within a sink can also establish communication with the cloud, creating an environment vulnerable to sniffing or theft of legitimate information. WBS could use Software as a Service (SaaS) cloud platforms, which provide sleek front-end and monitoring features, pushing significant jobs like management, communication, computation, and storage to trusted third parties, thereby raising \textbf{privacy} concerns \cite{Tintel2015}. \textbf{Deceptive and ambiguous privacy policies} trap users, allowing parties to sell information they collect to third party-managed databases which might not have strong information security protocols in place. \textbf{Session hijacking} attacks (cookie and session thefts, brute-force) that steal sensitive data flowing into the sink have also been studied \cite{Ching2016}. Some WBS applications in the sinks grant permissions to other apps for exchange or access to their information, which could pose a threat for BDI. \textbf{Weak bonding policies} between the sink and its apps could pose a significant threat, causing external \textbf{device mis-bonding} attack (DMB) \cite{Naveed2014}. Additionally, \textbf{JTAG-Read} (R) and boundary scan based attacks have been shown to access the memory of sinks, and allow adversaries to read the contents in its memory \cite{Rahmanm2016}. \textbf{Side-channel attacks} such as differential power attack that monitor the power consumption profile of the sink to steal secret keys have also been successfully conducted \cite{Ani2016,Maiti2015,Kocher1999,MArtin2004}.

Providing \textbf{trust-based management} for the transfer and storage of user-sensitive data is a viable addition to secure communication channels \cite{Linn2011,Manjusha2015}, as it limits data sharing to between WBAN sensors which have a valid trust in each other. With \textbf{trust revocation}, the system maintains a time-variant dynamic trust model, generating flags when one of the sensors has failed to establish trust with any other sensor in the same network. When one sensor is compromised, the trust model altered as a consequence would flag an alert and island that sensor to avoid further compromise \cite{Leeh2015}.
\textbf{Compartmentalization} of wearable data has proven to be a good countermeasure against thefts and frauds \cite{Someren2015}. Designing wearable applications to store vital information in segregated, encrypted, application-specific chunks restricts data replication. \textbf{Data protection} makes sure critical applications run un-preempted, data is securely backed up and recoverable, and allows administrators to map data and information flow between WBANs, sinks and cloud without jeopardizing integrity and privacy \cite{Weii2014,Duan2005}.
In defense to hijacking, BDI and JTAG-R, data communication and storage protocol using two \textbf{pseudo-random numbers} generated through symmetric and secret keys was proposed \cite{Rahmanc2014}, its implementation focused more on WBS where single-node WBAN and sink nodes are integrated into the same device. To counter DMB attacks, an OS-level protection method called \textbf{Dabinder} was proposed to enforce secure bonding policies whenever an app tried to establish Bluetooth connection or pair with the sink \cite{Naveed2014}.

\subsubsection{$PR$ Threats, Attacks and Defenses.}\label{sec:4.2.4} Most WBS today have feature extractor and matcher modules located in the same device: a separate sink, or wearable itself. Considering, WBS operate in noisy environments where other devices also communicate, signals are prone to noise and interference, which demand more processing from feature extractors and matchers \cite{Nisha2014}. This is in contrast with TBS that operate in controlled environments. More computation resources increase attack surface, paving way for \textbf{cryptanalytic attacks} that target pseudorandom number generators used by encryption methods in WBS \cite{Kelsey1998}. Acoustic key search, electromagnetic attacks, ciphertext, birthday, preimage, and key generation also come under this category.

Defending these attacks ranges from resetting sinks and erasing any trace of stored data, to more advanced methods that look at \textbf{match-level fusion} of signals to increase complexity of an attack to demotivate the adversary. Alternatively, extractor and matcher modules could be designed to function in frequency-domain. Since signals measured by wearables are time-variant, they could be directly correlated with owner's activities. Transforming signals into frequency-domain and using \textbf{spectral analyses} to operate on the features could make inference less explicit. The viability of \textbf{one-step two-factor authentication} scheme was discussed for wearable biosensors in the contexts of keystroke, EEG, hand geometry, and hand gesture \cite{JChuang2014}. \textbf{Case-based reasoning} could be used between extractor-matcher modules and other associated modules of the sinks.

\textbf{Key Takeaway Points:}  The following are the key takeaway points from this section:
\begin{enumerate}
\item The literature on threats, attacks and defenses for TBS and WBS can be modality ($M$) or Technology ($T$) attacks, where $T$ attacks can be System ($Sys$), Pattern Recognition ($PR$) or Hybrid ($H$) attacks for TBS, and $Sys$ could be $Sys$-$WBAN$ and $Sys$-$Sink$ for WBS
\item All the attacks discussed impact confidentiality, integrity and/or availability of the modality and/or template, and hence the privacy of users
\item Physical attacks are harder to conduct on WBS than on TBS since the former employ passive authentication and are more complexly networked together
\item Systemic security measures such as adversarial machine learning and game theory could be used to defensively learn adversary strategies and psychology before proactively resolving the attacks
\item Most of the attacks that target TBS and WBS tend to indirectly affect their performance, especially factors like energy, network, and spectral efficiencies, and throughput/goodput, which establishes a strong coupling between security concerns and performance
\item Modality characteristics determine the impact of $M$ attacks while performance metrics determine that of $T$ attacks. Security of TBS, on the other hand, depends on modality, sample measured, template, and feature matching/decision-making
\end{enumerate}

\section{Factors Contributing to WBS Design Solutions}
\label{sec:6}
Despite numerous WBS products already in the market and many more emerging, the design considerations for WBS is still an evolving subject of research. Many principles impacting WBS design have been identified:\cite{shaun,bigdwear,bigdhealth,ergonomic}
\begin{itemize}
\item Human interaction with WBS could be in four ways: audio, visual, tactile and haptic \cite{humanfactors}. These interaction modes, which form a fundamental aspect of WBS design solutions, are influenced by different parameters like cognitive ease, cognitive overload, intuitiveness, comprehension and perception
\item With the WBS required to be always online and support passive authentication and/or identification, the sensors generate ample amount of data over a given period of time. The WBS design must account for not just data acquisition, but appropriate data management- from sanitation, processing and storage to destruction. Some of the evaluation metric factors discussed earlier like accuracy, energy and network efficiencies, will play key roles in shaping WBS design with respect to its data handling requirements \cite{bigdwear}
\item Aesthetics have always been at the core of WBS, with designs catering to elegant, fashion-conscious models that are powerful as well as easily wearable \cite{ergonomic}. It is a significant consideration, since a good design for WBS is a proper balance between the technology's intelligence and its appearance and user-friendliness
\item The different WBS modality characteristics surveyed in Section \ref{sec:3} impact WBS design. This is so, because every modality has its own properties (such as uniqueness, universality, permanence, acceptability and robustness against circumvention), requirements for measurement (such as collectibility, susceptibility to interference from external signals and ambient noise, and invasiveness) and constraints for processing and maintenance (such as security, reliability, mobility and variability). The heterogeneity between different modalities requires WBS design to be tweaked accordingly. For example, a WBS designed to work on EEG might not be a good design for EMG or PPG
\item As detailed in Section \ref{sec:3.2}, different factors that impact WBS performance also contribute to their design considerations. Physiology, number and placement of sensors, environmental elements, and sensory heterogeneity affect the criteria for WBS designs
\item Performance of WBS can be impacted by attacks, which affect security design aspects:
\begin{enumerate}
\item Attacks that delay or corrupt the data packets within WBAN or WBAN-Sink communication could affect throughput and goodput, an unchecked manipulation of which could result in faulted WBS that could even be life-threatening
\item The $M$ class attacks have a strong coupling with the modality characteristics like reliability, security and circumvention. They impact confidentiality and integrity of sensitive data. Considering revoking biometrics is impossible (unless revocable technology is fused with the system), a successful compromise can impact the system performance and operations adversely
\item DoS and other attacks that target the availability of data could drain the battery of WBS or trigger excessive usage of critical resources like network bandwidth, causing WBS to undergo preemptive shutdown that, in certain use-cases, could not only be disruptive to system operation but also be fatal to the owner
\item It is known that WBS are sensitive to external noises, considering their energy and spectral efficiencies depend on them. However, jamming attacks distort the legitimate signals by adding noise to impact WBS performance
\end{enumerate}
\item Many recent works have proposed novel WBS materials: stretchable silicon-based elastomers like polydimethylsiloxane or Ecoflex for substrates and metal-based conductors for electrodes \cite{materials}. They exhibit properties like biocompatibility, stretchability, conductivity, malleability, ability to withstand strain and stress, and have a low failure rate
\end{itemize}

\section{Open Research Challenges and Future Directions}
\label{sec:7}
WBS have proven their mettle in healthcare and fitness, but their security concerns in these domains as well as when considered solely for authentication and identification purposes have significant rooms for improvement \cite{Darwish2011,Katrin2015,Lymberis2003,Williams2016,Samaher2017}. Specifically, key open research areas in WBS are:
\begin{itemize}
\item Evolving security scape: The increased adoption of IoE principles into WBS will continue to push boundaries in terms of signal processing, connectivity, data processing and management, measurement accuracy, convenience and aesthetics, but the constantly evolving nature of attacks will pose a significant and persistent threat to their tech-scape. Recently, the use of blockchain technology to store and analyze WBS data was explored to offer a personalized healthcare for customers~\cite{surrey2017}. The approach used distributed ledger technology and machine learning to store and access data of users in a secure manner. More work is foreseen in areas that combine data analytics, cyber-physical security, and biometrics
\item Changes to business models: With cloud and edge-based decentralized data analytics becoming the norm for WBS, business models and intellectual property must be redefined to adapt and support the relevant advancements in this newly redefined environment. WBS are viewed as one of the first technologies that pave way for a customer-driven market where choices of end-consumers drive the industries
\item WBS Big Data: With increasing adoption of WBS by consumers, the number of sensors collectively generating data will increase. Considering a single user can have a WBAN of multiple sensors that are always online and ubiquitously churn new data periodically, significant advances to manage, process and analyze the wealth of raw information in a decentralized manner must be developed. The data thus generated will find multiple uses ranging from consumer analytics to business intelligence and personalization of services
\item Newer methods to leverage the power of emerging modalities such as voice, signature, iris  and human interactions have recently been explored. Hand-worn devices such as smartwatches and fitness wear have been used to verify signatures and prevent frauds in the financial sector. A study used voice recognition to perform two-factor authentication on WBS~\cite{cburt2017}. This technology generates speech embedded with a random code that a browser then plays. The signals are then captured by the WBS to perform authentication. Google recently patented an iris scanning contact lens that uses the light reflected by the iris to perform authentication~\cite{gpatent}. These technologies increase the likelihood that the traditionally used modalities will now be exploited to solve the emerging challenges of security and also not jeopardize user privacy and invasiveness
\item Revision to policies: With changes in business models, policymaking follows. With consumers demanding greater transparency, end-to-end analytics, peer-to-peer information exchange, localized privacy-aware processing, and much more, restructuring of policies and legislation will be inevitable
\item Social behavior and acceptance: WBS are still viewed as invasive technologies by a majority of consumers in the market. Hence, besides revisions to policies, social analytics and acceptance testing through controlled experiments, surveys, interviews, awareness and outreach must be conducted by the companies manufacturing WBS, thereby creating a pipeline to gather consumer information from WBS data for offering personalized services
\end{itemize}

\vspace{-0.5cm}
\subsection{WBS Datasets and Access}
To further the future research in WBS, data acquisition from different biometric modalities is important. However, setting up devices, recruiting the right mix of participants, and gathering measurements is not an easy task. This is where online data repositories come into importance. These repositories include, but are not limited to, the UCI's Machine Learning Repository, Kaggle, knowledge discovery in datasets, KDNuggets, Wiley online repository, and even research repositories such as the one hosted by the Biometrics Security Lab for the PPG datasets~\cite{uciml2017,biosecppg,elena2018}.

Although medical professionals have been collecting biometric data over years through clinical trials, experiments, patient examinations and observations, the WBS have increased the rate of data acquisition and the types of data being collected. This opens emerging challenges such as data mining, minimizing signal-to-noise ratio, determining the length and persistence of data collection and storage, respectively, for effective analysis and diagnosis of associated health risks in case of fitness applications and being adaptive to potential consumer preference changes in case of applications of leisure~\cite{deloittereport}. The lack of proper standardization to ensure data consistency, trust, and integrity in interpretation and analysis is a challenge to gain insights from WBS datasets.

One of the key research gaps identified in~\cite{novettareport} is obtaining intelligence for diagnosis from raw WBS data without the involvement of human agents to manually parse and contextualize. Legal, ethical, administrative and technical concerns have been identified as barriers to widespread use of third party data collected by the above sources. The report also identifies the NIST Biometric \& Forensic Research Database Catalog that serves as a central repository for publicly available biometric and forensic datasets~\cite{nistrepo}. Collaborative efforts by the U.S. Military Academy and the Defense Advanced Research Projects Agency (DARPA) have also enabled the process of biometric data measurement and collection, followed by dissemination. Wearable data acquisition and/or access is still an emerging research problem with significant administrative, ethical, legal, and technical implications that must be sorted out.


\section{Conclusion}
\label{sec:8}
One of the important underpinning inferences visible from the emerging research in the area of WBS is that their operational dynamics are different from those of TBS which have been available in the market for a long time. Although the research community is aware of the key differences between the two systems, the factors that contribute to these differences are not well researched, summarized or discussed. To bridge this gap among the recent works of literature, this paper conducts a comprehensive survey on three distinct but interdependent aspects of biometric systems: the characteristics of modalities they use, the metrics used to evaluate their system performance, and their security and privacy. Initially, to help appreciate the differences between TBS and WBS, the paper reviews and contrasts the above three aspects for both types of biometric systems with equal emphasis. However, given the future research is geared more towards the security and privacy concerns of WBS, the paper highlights how the design solutions to enhance security and privacy are impacted by WBS modality characteristics and performance factors. Thereby, the survey is aimed at not only summarizing, but also using the surveyed results to contribute to the literature a clear understanding of the differences between TBS and WBS, advancements in WBS technology and research, and the factors that impact their security and privacy design. It is inferred that most modalities considered invasive in TBS could be powerful in the wearable environment. WBS require additional metrics to measure their performance owing to their ubiquitous nature, and the landscape of security is often different from that of TBS. In the end, open research challenges and future directions for research in biometric systems was briefly discussed.


\bibliographystyle{ACM-Reference-Format}
\bibliography{sample-bibliography}